 \documentclass[final,5p,times,twocolumn,authoryear]{elsarticle}

\usepackage[authoryear]{natbib}
\usepackage{graphicx}
\usepackage{subcaption}
\usepackage{amssymb,amsmath}
\usepackage{pdflscape}

\usepackage{lipsum}
\usepackage{epsfig}
\usepackage{epstopdf}
\usepackage[utf8]{inputenc}
\usepackage{newunicodechar}
\usepackage{textgreek}
\usepackage{physics}
\usepackage{inputenc}

\usepackage{afterpage}

\usepackage{amsthm}

\usepackage{multirow}
\usepackage{array}
\usepackage{url} 
\usepackage{xcolor}
\usepackage{adjustbox}
\usepackage{soul}
\usepackage{orcidlink}
\usepackage{multirow}
\usepackage{array}

\usepackage{longtable}
\usepackage{threeparttable}

\journal{High Energy Astrophysics}

\begin{document}

\begin{frontmatter}



\title{Exploring Year-timescale Gamma-ray Quasi-Periodic Oscillations in Blazars: Evidence for Supermassive Binary Black Holes Scenario}







\author[first]{Ajay Sharma\,\orcidlink{0000-0002-5221-0822}}
\ead{ajjjkhoj@gmail.com}

\author[first]{Sakshi Chaudhary}
\ead{gurjarsakshi51@gmail.com}

\author[second]{Aishwarya Sarath}
\ead{aishwarya.sarath@mail.udp.cl}

\author[third]{Debanjan Bose\,\orcidlink{0000-0003-1071-5854}}
\ead{debanjan.tifr@gmail.com}


\affiliation[first]{{S N Bose National Centre for Basic Sciences,},addressline={Block JD, Salt Lake}, city={Kolkata},
           postcode={700106}, 
            state={West Bengal},
            country={India}}

\affiliation[second]{{Instituto de Estudios Astrofísicos, Facultad de Ingeniería y Ciencias, Universidad Diego Portales,},addressline={Av. Ejército Libertador 441}, city={Santiago}, country={Chile}}

\affiliation[third]{{Department of Physics, Central University of Kashmir,}, 
            addressline={Ganderbal}, 
            postcode={191131}, 
            state={Kashmir},
            country={India}} 

\cortext[cor1]{Corresponding author: Ajay Sharma, Debanjan Bose }


\begin{abstract}

A comprehensive analysis of quasi-periodic oscillations (QPOs) in the gamma-ray emissions of blazars. Utilizing 15 years of Fermi-LAT observations of seven blazars in our sample, we identify both long-term and transient quasi-periodic oscillations in the gamma-ray light curves, with timescales ranging from a few months to years. These periodicities were detected using the Lomb-Scargle periodogram and REDFIT techniques. To robustly evaluate the statistical significance of the quasi-periodic signals observed in the Lomb-Scargle Periodograms, 30,000 synthetic $\gamma$-ray light curves were generated for each source using a stochastic model known as the Damped Random Walk (DRW) process. To investigate the physical origin of the observed gamma-ray QPOs with different timescales, we explore several plausible scenarios, with particular emphasis on a relativistic jet hosted by one of the black holes in a supermassive binary black hole system, jet precession, and helical motion of magnetized plasma blob within the jet. The $\gamma$-ray light curves exhibiting long-timescale quasi-periodic oscillations (QPOs) are analyzed within the framework of a supermassive binary black hole (SMBBH) model, employing a Markov Chain Monte Carlo (MCMC) approach, allowing us to constrain key physical parameters such as the jet Lorentz factor ($\Gamma$) and the viewing angle between the observer's line of sight ($\psi$) relative to the spin axis of SMBH.

\end{abstract} 



\end{frontmatter}




\section{Introduction}
\label{introduction}

Active galactic nuclei (AGNs) are among the most energetic astrophysical sources in the universe, powered by the accretion of matter from their host galaxies onto supermassive black holes (SMBHs) with masses ranging from $10^6$ to $10^{10} \ \rm{M}_\odot$ \citep{soƚtan1982masses}. A particularly extreme class of AGNs, known as blazars, are radio-loud sources distinguished by their exceptional luminosities, with bolometric outputs spanning $10^{41}$ to $10^{48}$ erg s$^{-1}$. These objects are characterized by powerful relativistic jets that are closely aligned with our line of sight (typically within $<5^\circ$) \citep{ghisellini1993relativistic, urry1995unified, blandford2019relativistic}, resulting in strong Doppler boosting of their emission. Blazars radiate across the entire electromagnetic spectrum, from radio wavelengths to very high energy (VHE; $>$100 GeV) $\gamma$-rays \citep{urry1995unified, ulrich1997variability, padovani2017active}. Blazars are generally divided into two subclasses based on their optical spectral features: BL Lacertae objects (BL Lacs) and flat-spectrum radio quasars (FSRQs). BL Lacs display nearly featureless, nonthermal optical spectra with weak or absent emission lines, whereas FSRQs exhibit prominent broad emission lines with rest-frame equivalent widths (EW) greater than 5  $\text{\AA}$ \citep{giommi2012simplified}.\par

Observational studies have revealed that blazars exhibit rapid and large-amplitude flux variations across the entire electromagnetic (EM) spectrum, from radio wavelengths to very high energy (VHE) $\gamma$-rays. These variations arise from jet-dominated nonthermal emission, which produces a characteristic double-humped spectral energy distribution (SED) \citep{ulrich1997variability, fossati1998unifying}. The observed variability provides critical insights into the internal structure of blazars, the underlying emission mechanisms, and the physical properties of their central supermassive black holes (SMBHs) \citep{ulrich1997variability, gupta2017multi}. Flux modulations in blazars occurs over a wide range of timescales, from just a few minutes to several years. Given that the central regions of blazars are extremely compact and challenging to resolve directly with current observational capabilities, analyzing short-timescale variability (on the order of minutes to hours) offers a powerful method to constrain the size of the emission regions. Moreover, by combining simultaneous multiwavelength observations with theoretical modeling, it is possible to place stringent constraints on the physical parameters of the relativistic jets \citep{blandford1982reverberation, tavecchio1998constraints, li2018fast, pandey2022detection}.\par

The flux variability in blazars is predominantly stochastic, nonlinear, and aperiodic, often well described by the Continuous Autoregressive Moving Average [CARMA(p,q)] model, also known as the red noise model \citep{kelly2009variations}. However, a small fraction of blazars exhibit quasi-periodic oscillations (QPOs), a rare phenomenon in AGNs characterized by regular, repeating patterns in their light curves. QPOs have been observed across the electromagnetic (EM) spectrum, with timescales ranging from minutes to years \citep{urry1993multiwavelength, wagner1995intraday, petry2000multiwavelength, sandrinelli2014long, gupta2019characterizing, mao2024radio}.\par

The diversity in QPO timescales points to a range of underlying physical mechanisms. Intraday QPOs (minutes to hours) may arise from rotating helical jet structures or kink instabilities \citep{raiteri2021dual, jorstad2022rapid}, while short-term QPOs (days to months) are often linked to helical motion of magnetized plasma blobs or accretion disk instabilities near the innermost stable circular orbit \citep{zhou201834, prince2023quasi}. Long-term QPOs (months to years) are commonly associated with supermassive binary black holes (SMBBHs) or large-scale jet dynamics \citep{begelman1980massive, graham2015possible, li2023quasi}. Notable candidates such as PG 1553+113 and OJ 287 have been proposed to host SMBBHs \citep{sillanpaa1988oj, ackermann2015multiwavelength}. In addition, other geometrical models — such as jet precession, Lense-Thirring precession, and pulsating accretion flows, have been proposed to explain observed QPOs \citep{romero2000beaming, rieger2005helical, stella1997lense}. Moreover, transient QPOs have been attributed to phenomena such as orbiting hotspots near SMBHs, magnetic reconnection, and helical motion of plasma blobs in magnetized jets \citep{gupta2008periodic, huang2013magnetic, banerjee2023detection, prince2023quasi, sharma2024detection, tantry2025study, sharma2025searching}. Overall, QPO studies offer a powerful tool for probing the physical conditions in blazar jets and the central engine of AGNs.

The paper is arranged as follows: Section \ref{sec:fermilat} covers Fermi-LAT observations of sources in our sample, including PKS 1424-41, PKS 0736+01, S2 0109+22, PKS 0244-470, PKS 0405-385, PKS 0208-512, and PKS 0035-252, and reduction techniques. In Section \ref{sec:QPO_search}, we search for quasi-periodic signals in $\gamma$-ray light curves of blazars in our sample by employing various techniques, including the Lomb-Scargle Periodogram (LSP) and REDFIT. Section \ref{sec:LC_modeling} explores the various plausible scenarios explaining the QPOs in $\gamma$-ray emission. Section \ref{sec:disscusion} presents the discussion and conclusion of this study.

\begin{table*}
\setlength{\extrarowheight}{7pt}
\setlength{\tabcolsep}{2pt}
\centering
\begin{threeparttable}
\caption{Summary of the $\gamma$-ray QPOs of individual blazars as reported in the literature.}
\label{tab:source_sample}
\begin{tabular}{c c c c c c c c c c c c}
\hline
\hline
Source  & 4FGL association & R.A. & Decl. & Redshift & Class & log $M_{BH}$  & QPO timescale & Cycle & \multicolumn{3}{c}{Reference}  \\
\cline{10-12}
 & & & & & & & & & $z$ & $M_{BH}$ & $P_{QPO}$\\
 & & & & ($z$) & & units of $(M_{\odot})$ &  ($P_{QPO}$)  [d] & & &  & \\
(1) & (2) & (3) & (4) & (5) & (6) & (7) & (8) & (9) & (10) & (11) & (12)\\
[+2pt]
\hline
PKS 0736+01 & J0739.2+0137 &  114.820 & 1.622 & 0.189 & FSRQ & 7.96 & - - & - - & (a) & (b) & - - \\ 
PKS 1424-41 & J1427.9-4206 & 216.98 & -42.10 & 1.522 & FSRQ & 9.65 & $\sim$341 & 6 & (c) & (d) & (c)\\
S2 0109+22 & J0112.1+2245 & 18.02 & 22.75 & 0.36 & BL Lac & 8.6 & $\sim$600 & 5.6 & (f) & (e) & (e)\\
PKS 0244-470 & J0245.9-4650 & 41.49 & -46.84 & 1.385 & FSRQ & 8.48, 8.32 & $\sim$225 & 8 & (h) & (g) & (h) \\
PKS 0405-385 & J0407.0-3826 & 61.76 & -38.43 & 1.285 & FSRQ & 8.7 & $\sim$1037 & 5 & (i) & (j) & (j)\\
PKS 0208-512 & J0210.7-5101 & 32.69 & -51.02 & 1.003 & FSRQ & 8.84 & $\sim$985 & $\sim$4 & (l) & (l) & (k) \\
PKS 0035-252 & J0038.2-2459 & 9.56 & -24.99 & 0.49 & FSRQ & 7.18 & - - &  - - & (b) & (b) & - -   \\
[+5pt]
\hline
\end{tabular}

\begin{tablenotes} 
\small
\item Note: The general information of blazars in our sample. Column (1): source name; Column (2): source name in the Fermi-LAT $4^{th}$ catalog; Column (3 - 4): coordinate of source; Column (5): redshift; Column (6): blazar type; Column (7): black hole mass; Column (8): reported QPO timescale in $\gamma$-rays; Column (9): number of QPO cycle; Column (10-12): references -- (a) \citep{abdalla2020hess}, (b) \citep{pei2022estimation}, (c) \citep{chen2024transient}, (d) \citep{fan2004black}, (e) \citep{zhang2023detection}, (f) \citep{magic2018broad}, (g) \citep{shaw2012spectroscopy}, (h) \citep{das2023detection}, (i) \citep{kedziora1997pks}, (j) \citep{gong2022quasiperiodic}, (k) \citep{penil2020systematic}, (l) \citep{ghisellini2010general}

\end{tablenotes}
\end{threeparttable}
\end{table*}

\section{Fermi-LAT observation} \label{sec:fermilat}

The Fermi Gamma-ray Space Telescope, launched by NASA on June 11, 2008, onboard two instruments: the Large Area Telescope (LAT) and the Gamma-ray Burst Monitor (GBM). Together, they offer comprehensive coverage of the gamma-ray sky across a broad energy range, from a few keV up to 500 GeV. The Fermi-LAT, a pair-conversion gamma-ray detector, is designed to detect high-energy gamma rays in the 20 MeV to 500 GeV range. It features a wide field of view exceeding 2 steradians, enabling coverage of approximately 20$\%$ of the sky at any given moment. Since its launch, Fermi-LAT has performed full-sky surveys every three hours, delivering near-continuous monitoring of gamma-ray emissions from a wide variety of astrophysical sources, including active galactic nuclei, pulsars, and gamma-ray bursts \citep{atwood2009large}. This frequent and extensive sky coverage has made Fermi-LAT an essential instrument for high-energy astrophysics. \par

The Fermi-LAT observations for all sources in our sample—including PKS 1424-41 ($z = 1.52$), PKS 0736+01 ($z = 0.189$), S2 0109+22 ($z = 0.36$), PKS 0244-470 ($z = 1.385$), PKS 0405-385 ($z = 1.285$), PKS 0208-512 ($z = 1.003$), and PKS 0035-252 ($z = 0.49$)—span from August 5, 2008 (MJD 54683) to April 1, 2025 (MJD 60766). Further details for each source are provided in Table~\ref{tab:source_sample}. For data extraction, we selected an energy range of 0.1–300 GeV and employed Pass 8 event selection criteria, specifically evclass == 128 and evtype == 3, as recommended by the Fermi-LAT collaboration. Events were chosen from a circular region of interest (ROI) with a radius of $10^\circ$ centered on each source. The analysis of $\gamma$-rays was performed following the standard procedures for point-source analysis using the Fermi Science Tools package (v11r05p3), provided by the Fermi Science Support Center. To reduce contamination from Earth’s limb, we applied a zenith angle cut of $< 90^\circ$. High-quality data were ensured by filtering the good time intervals (GTIs) using the expression: \texttt{$\text{(DATA\_QUAL > 0) \&\& (LAT\_CONFIG == 1)}$}. We used \texttt{GTLTCUBE} and \texttt{GTEXPOSURE} tools to calculate the integrated livetime as a function of sky position and off-axis angle and exposure, respectively. For modeling the diffuse background, we employed the Galactic diffuse emission model  $\text{gll\_iem\_v07.fits}$\footnote{\label{fermi}\url{https://fermi.gsfc.nasa.gov/ssc/data/access/lat/BackgroundModels.html}} and the isotropic background model $\text{iso\_P8R3\_SOURCE\_V3\_v1.txt}$\footref{fermi}. The XML source model file, which includes source positions and spectral shapes, was generated using the \textit{make4FGLxml.py} script. We then performed an unbinned likelihood analysis using the \textit{GTLIKE} tool \citep{cash1979parameter, mattox1996likelihood}, adopting the instrumental response function (IRF) "$\text{P8R3\_SOURCE\_V3}$". To evaluate the detection significance of each source, we used the \textit{GTTSMAP} tool to compute the test statistics (TS), defined as TS = 2$\Delta$log(likelihood) = -2log($\frac{L}{L_0}$), where $L$ and $L_0$ are the maximum likelihood of the model with and without a point source at the target location and maximum likelihood value fitted by the background model, respectively. The significance of finding the source at the specified position is assessed by the TS value with TS $\sim \sigma^2$ \citep{mattox1996likelihood}.\par
\begin{figure*}
    \centering
    \includegraphics[width=0.95\textwidth]{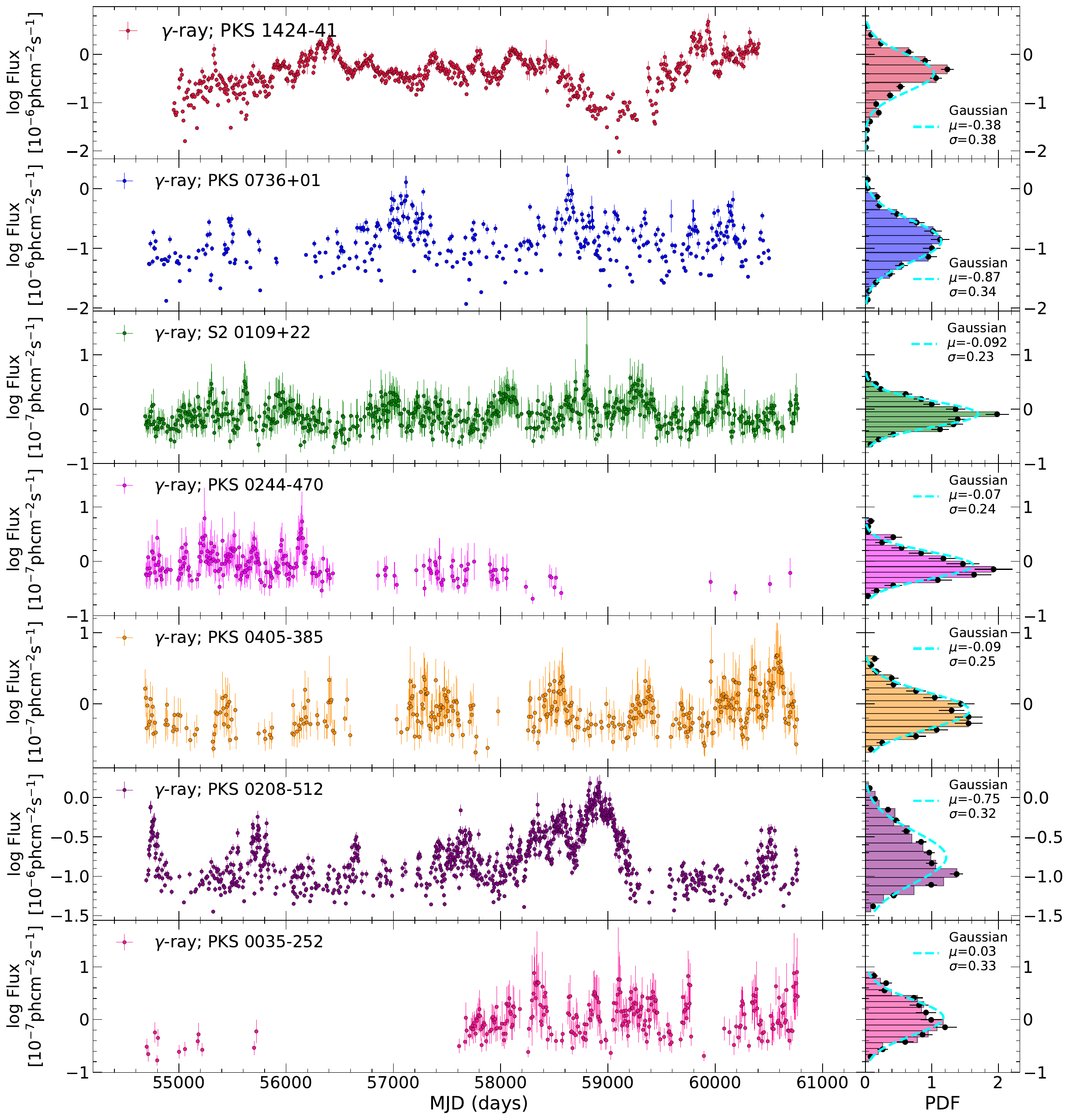} 
    \caption{Weekly binned $\gamma$-ray light curves of the blazars listed in Table~\ref{tab:source_sample} are shown in the left panel. The right panel shows the PDFs of logarithmic $\gamma$-ray flux values, Sect.~\ref{sec:flux_distribution}, with colors corresponding to the left panel. The dashed lines represent the Gaussian profiles fitted to these distributions.}
    \label{Fig-LC_FLUX_distribution}    
\end{figure*}

We adopted a criterion with TS($\ge$9) for data points in the light curves of each source, and weekly binned light curves of all blazars are generated using Fermipy\footnote{\url{https://fermipy.readthedocs.io/en/latest/}}. The resulting weekly binned $\gamma$-ray light curves are shown in Figure \ref{Fig-LC_FLUX_distribution}.


\section{\rm{Gamma-ray quasi-periodic oscillations and results}}\label{sec:QPO_search}
We adopted various methodologies in search of potential periodic signals in the $\gamma$-ray light curves of blazars in our sample. Figure~\ref{Fig-LC_FLUX_distribution} illustrates the weekly binned $\gamma$-ray light curves along with the flux distribution of log flux fitted with a normal distribution. The Lomb-Scargle periodogram (LSP) and REDFIT methods were used to analyze the light curves. A detailed description and the observed findings from all methods mentioned above are given in the following section~\ref{LSP}, \ref{redfit}.

\subsection{Lomb-Scargle Periodogram}\label{LSP}
The Lomb-Scargle periodogram (LSP) \citep{lomb1976least, scargle1979studies} is one of the most widely used approaches to identify any potential quasi-periodic signal in the $\gamma$-ray light curves. This approach is capable of handling unevenly sampled light curves efficiently by reducing the impact of noise and gaps and provide a precise measurement of the identified periodicity. The analysis incorporated the GLSP package to compute the Lomb-Scargle (LS) power. The mathematical expression of LS power is given as \citep{vanderplas2018understanding}:

\begin{equation}
\begin{split}
P_{LS}(f) = \frac{1}{2} \bigg[ &\frac{\left(\sum_{i=1}^{N} x_i \cos(2\pi f (t_i - \tau))\right)^2}{\sum_{i=1}^{N} \cos^2(2\pi f (t_i - \tau))} \\
&+ \frac{\left(\sum_{i=1}^N x_i \sin(2\pi f (t_i - \tau))\right)^2}{\sum_{i=1}^N \sin^2(2\pi f (t_i - \tau))} \bigg]
\end{split}
\end{equation}

where, $\tau$ is 
\begin{equation}
    \tau = \frac{1}{4\pi f}\rm{tan^{-1}} \left(\frac{\sum_{i=1}^N sin\left(4\pi f (t_i ) \right)}{\sum_{i=1}^N cos\left(4\pi f (t_i) \right)} \right)
\end{equation}

where, we selected the minimum frequency $\left( f_{min} \right)$ and maximum frequency $\left( f_{max} \right)$ in temporal frequency range as 1/T and 1/2$\Delta T$, respectively, and here T and $\Delta T$ represent the total observation time frame and the time difference between two consecutive points in the light curve, respectively.\par
  
\begin{figure*}
    \centering
    \includegraphics[width=0.44\textwidth]{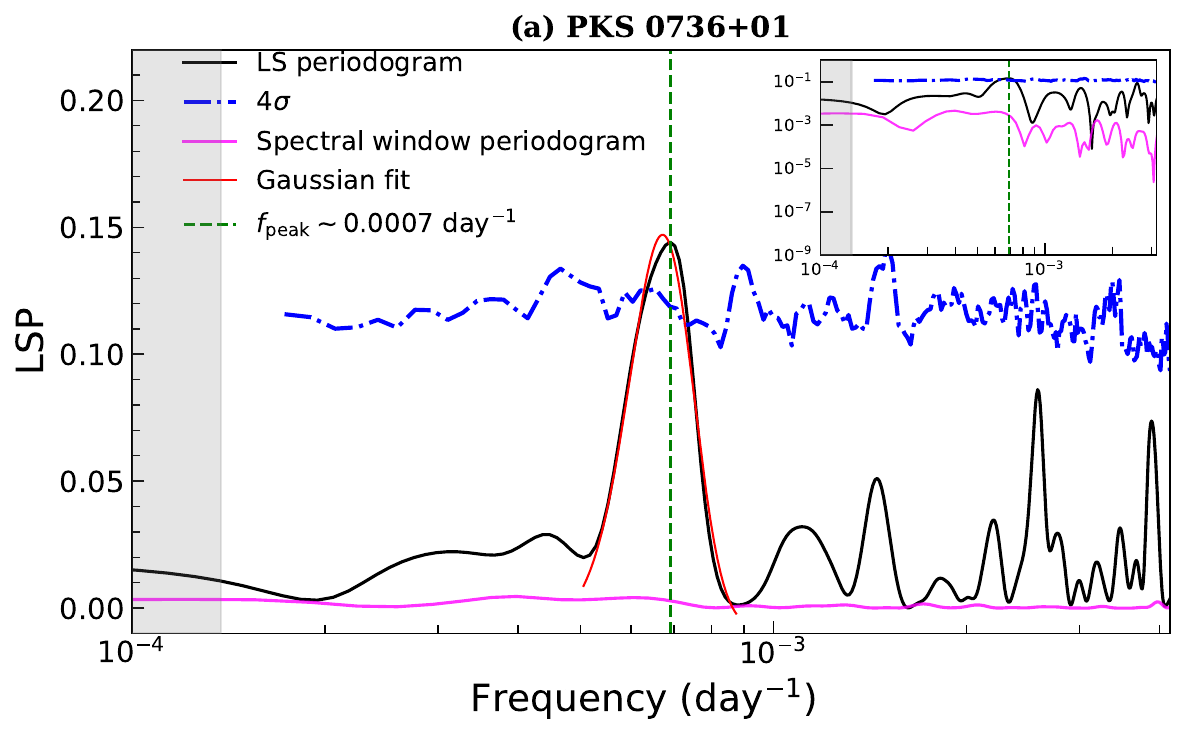}\hspace{1pt}
    \includegraphics[width=0.44\textwidth]{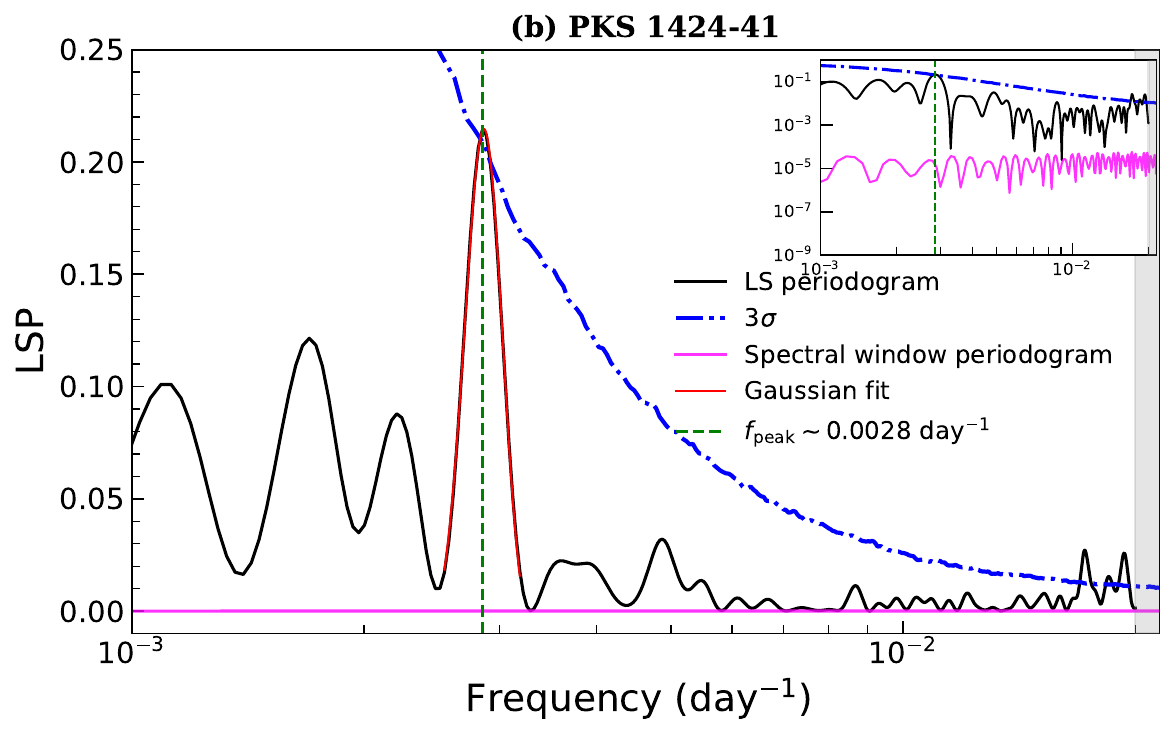}
    \vspace{1pt}
    \includegraphics[width=0.44\textwidth]{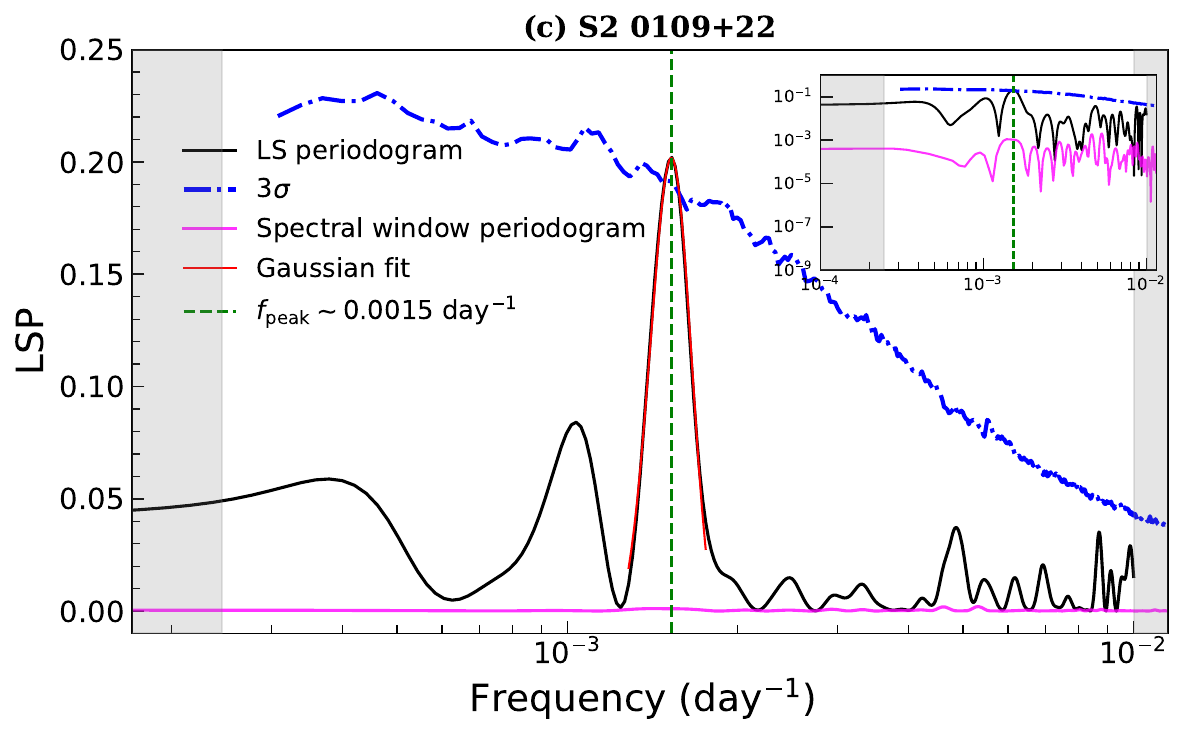}\hspace{1pt}
    \includegraphics[width=0.44\textwidth]{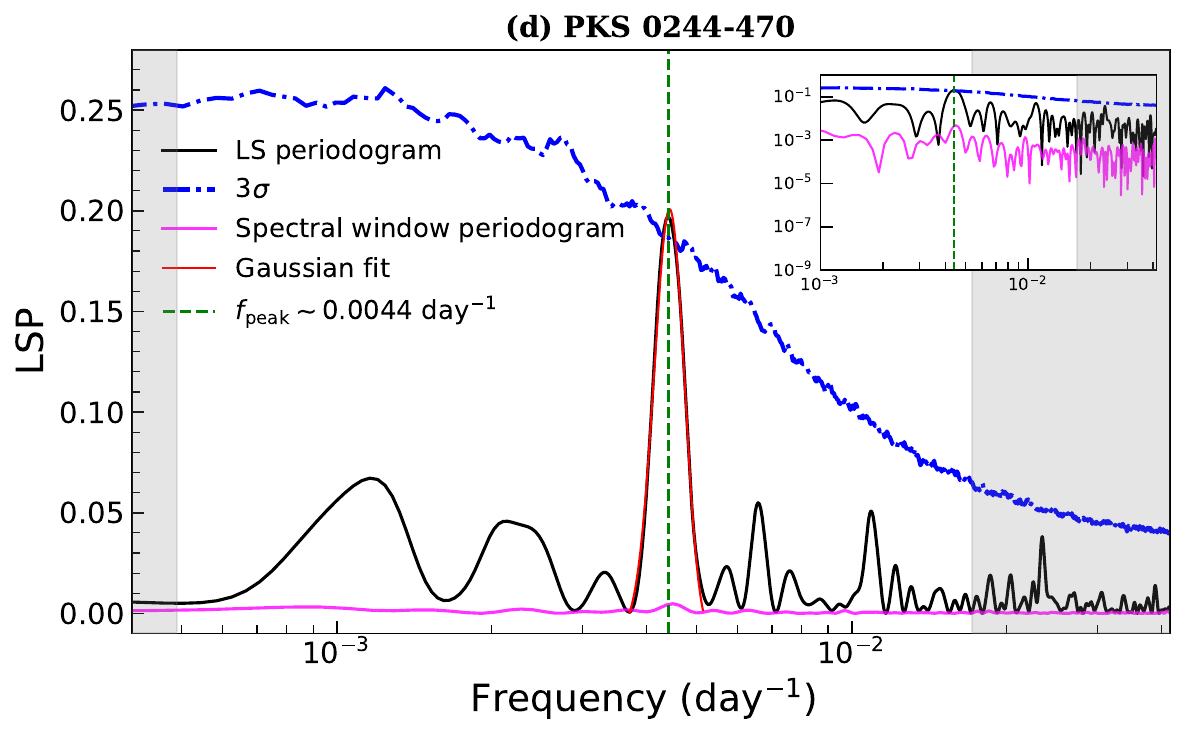}
    \vspace{1pt}
    \includegraphics[width=0.44\textwidth]{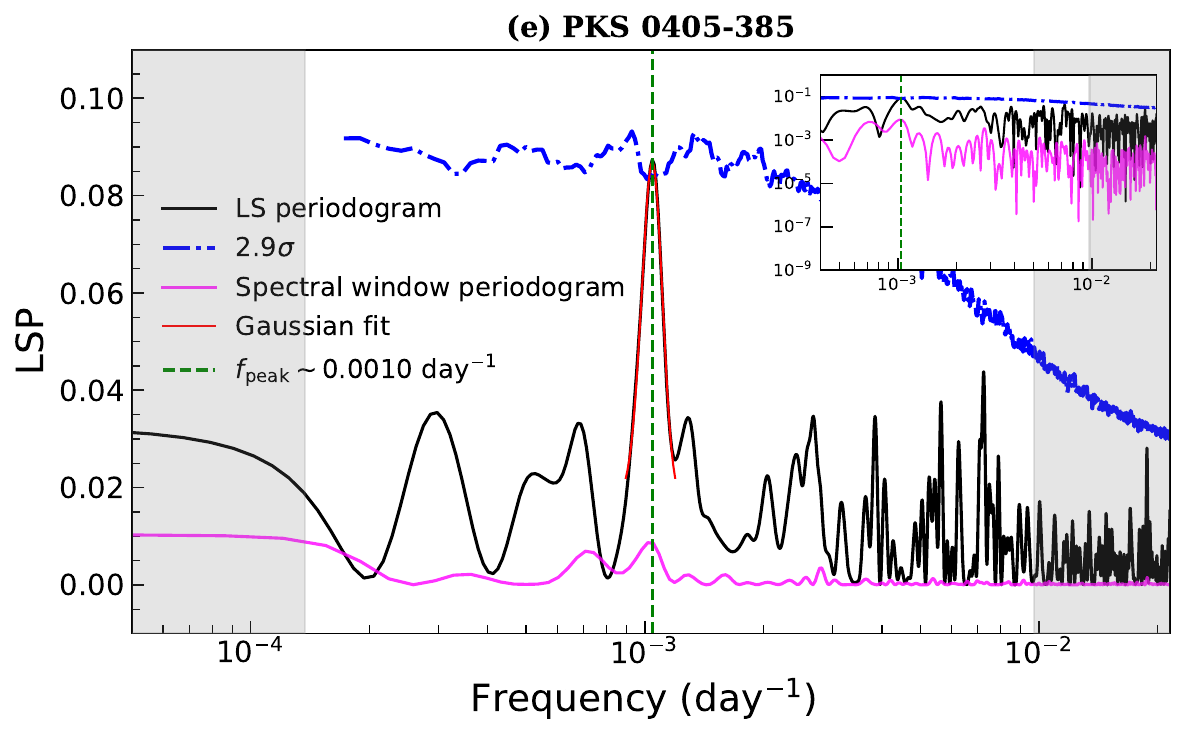} \hspace{1pt} \includegraphics[width=0.44\textwidth]{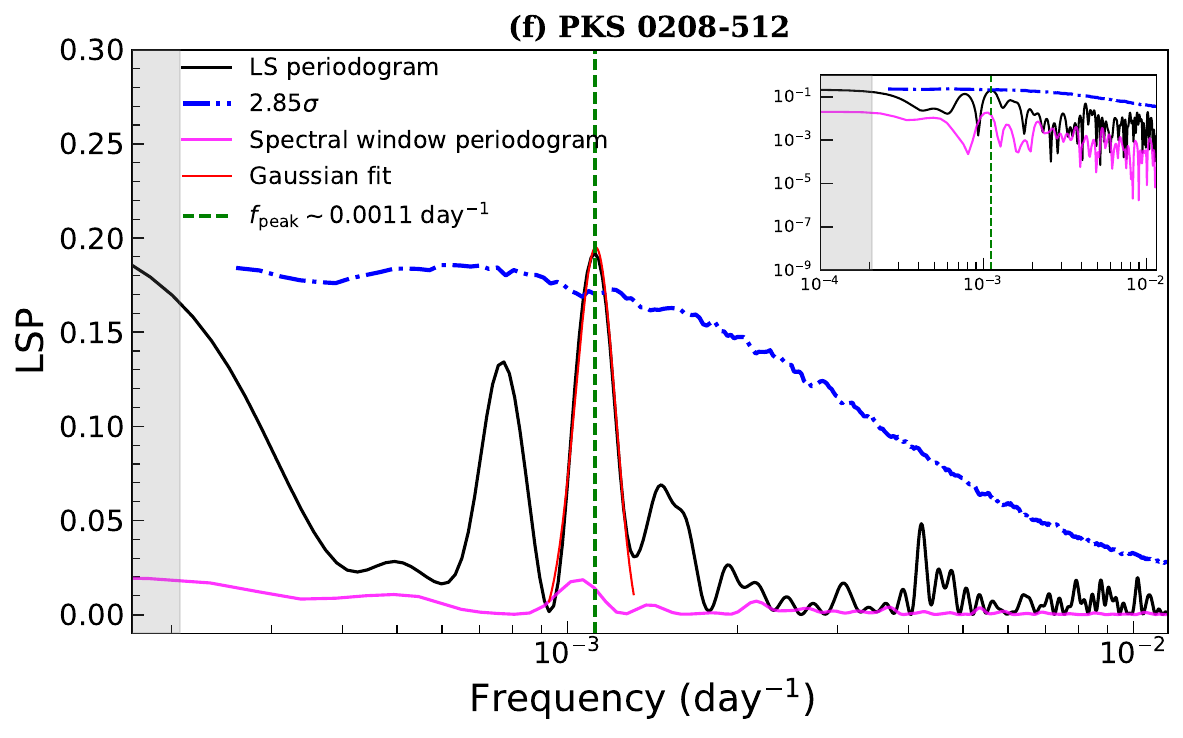}
    \vspace{1pt}
    \includegraphics[width=0.44\textwidth]{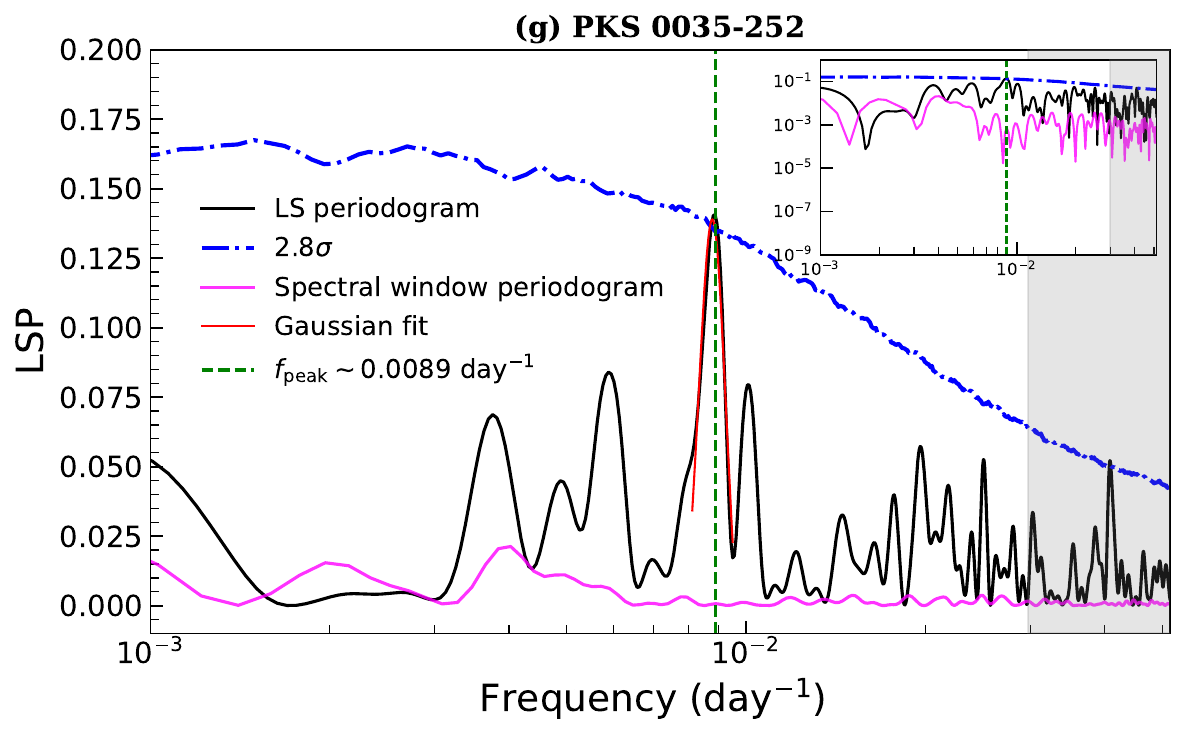}
    \caption{Lomb-Scargle periodograms of the $\gamma$-ray light curves for the blazars listed in Table~\ref{tab:source_sample}. The black curves represent the periodograms of the observed light curves, while the pink curves show the corresponding spectral window functions. The blue horizontal lines represent the local significance thresholds in the Lomb-Scargle periodograms: $4\sigma$ for PKS 0736+01, $3\sigma$ for PKS 1424-41, S2 0109+22, and PKS 0244-470, $2.9\sigma$ for PKS 0405-385, $2.85\sigma$ for PKS 0208-512, and $2.8\sigma$ for PKS 0035-252, estimated from 30,000 DRW-based synthetic light curves. Red curves denote Gaussian fits to the dominant peaks in the periodograms. The vertical dotted green line represents the position of the dominant peak. Shaded regions mark the unreliable frequency ranges, determined based on the mean cadence and temporal baseline of the light curves. For better clarity, insets display the logarithmic versions of the periodograms.}
    \label{Fig-DRW_LSP}    
\end{figure*}

The Lomb-Scargle Periodogram (LSP) analysis identified prominent QPO features in the $\gamma$-ray light curves of blazars, with the corresponding oscillation frequencies summarized as follows: PKS 0736+01 exhibits a QPO frequency of $(0.69 \pm 0.091) \times 10^{-3} \ \mathrm{day^{-1}}$, PKS 1424-41 shows $(0.28 \pm 0.019) \times 10^{-2} \ \mathrm{day^{-1}}$, S2 0109+22 has $(0.15 \pm 0.015) \times 10^{-2} \ \mathrm{day^{-1}}$, PKS 0244-470 displays $(0.44 \pm 0.035) \times 10^{-2} \ \mathrm{day^{-1}}$, PKS 0405-385 shows $(0.10 \pm 0.0074) \times 10^{-2} \ \mathrm{day^{-1}}$, PKS 0208-512 has $(0.11 \pm 0.010) \times 10^{-2} \ \mathrm{day^{-1}}$, and PKS 0035-252 shows $(0.28 \pm 0.019) \times 10^{-2} \ \mathrm{day^{-1}}$. The uncertainties on the detected frequencies were estimated by fitting the QPO peaks with Gaussian profiles and adopting the half-width at half-maximum (HWHM) as the frequency error. The findings are shown and tabulated in Figure~\ref{Fig-DRW_LSP} and Table~\ref{tab:QPO}, respectively.  

\subsection{\rm{REDFIT}}\label{redfit}
The light curves of AGNs are typically unevenly sampled, of finite duration, and predominantly influenced by red noise, which arises from stochastic processes occurring in the accretion disk or jet. Red noise spectrum are characteristic of autoregressive processes, where current emission is related to past emission. The emissions from AGNs are effectively modeled using a first-order autoregressive (AR1) process. The software programme \textsc{REDFIT}, developed by \citep{schulz2002redfit}, is specifically designed to analyze the stochastic nature of AGNs dominated by red noise. This software fits the light curve with AR(1) process, where the current emission ($r_t$) depends linearly on the previous emission ($r_{t - 1}$) and a random error term ($\epsilon_t$). The AR(1) process is defined as:

\begin{equation}
    r(t_i)=A_i r(t_{i-1}) + \epsilon(t_i)
\end{equation}

where $r(t_i)$ is the flux value at time $t_i$ and $A_i = exp\left( \left[ \frac{t_{i-1} - t_i}{\tau}\right] \right) \in [0,1]$, A is the average autocorrelation coefficient computed from mean of the sampling intervals, $\tau$ is the time-scale of autoregressive process, and $\epsilon$ is a Gaussian-distributed random variable with zero mean and variance of unit. The power spectrum corresponding to the AR(1) process is given by 

\begin{equation}
    G_{rr}(f_i) = G_0 \frac{1 - A^2}{1 - 2 A cos\left( \frac{\pi f_i}{f_{Nyq}} \right) + A^2}
\end{equation}

where $G_0$ is the average spectral amplitude, $f_i$ are the frequencies, and $f_{Nyq}$ is representing the Nyquist frequency. \par
In our study, we used the publicly available 
\textsc{REDFIT}\footnote{\url{https://rdrr.io/cran/dplR/man/redfit.html}} code to analyze the light curve.

\begin{figure*}
    \centering
    \includegraphics[width=0.47\textwidth]{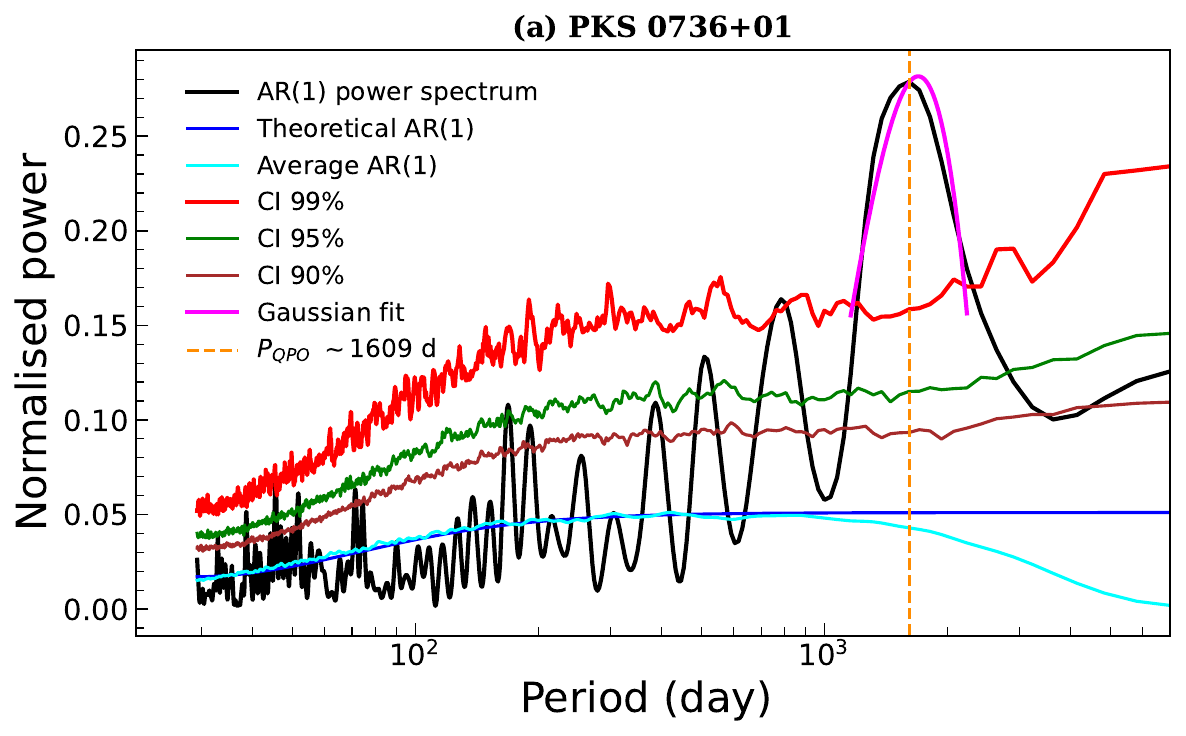} \hspace{1pt}
    \includegraphics[width=0.47\textwidth]{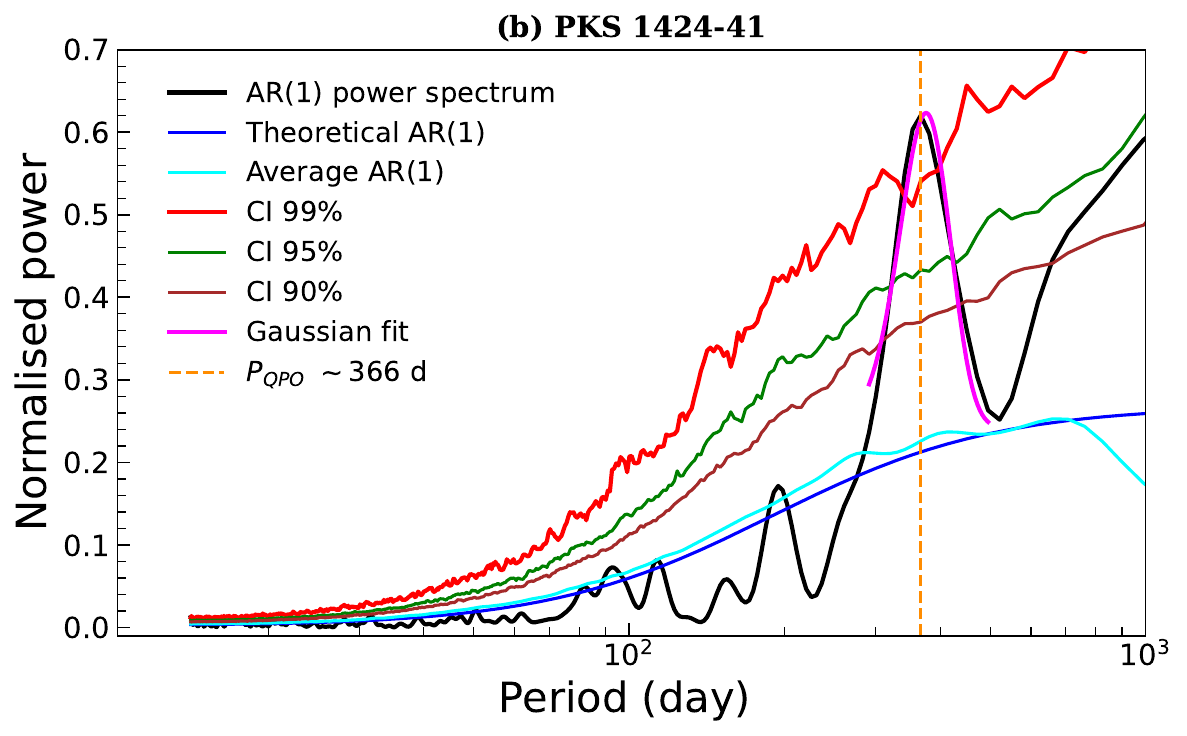}
    \vspace{1pt}
    \includegraphics[width=0.47\textwidth]{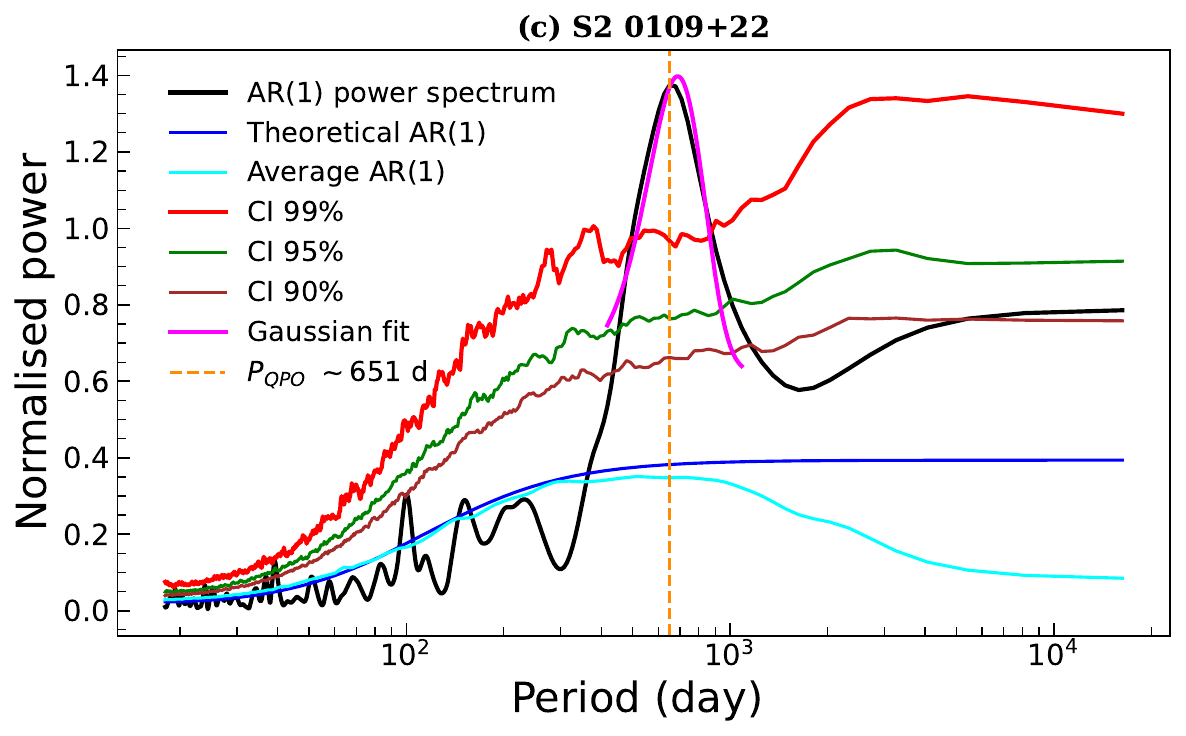}\hspace{1pt}
    \includegraphics[width=0.47\textwidth]{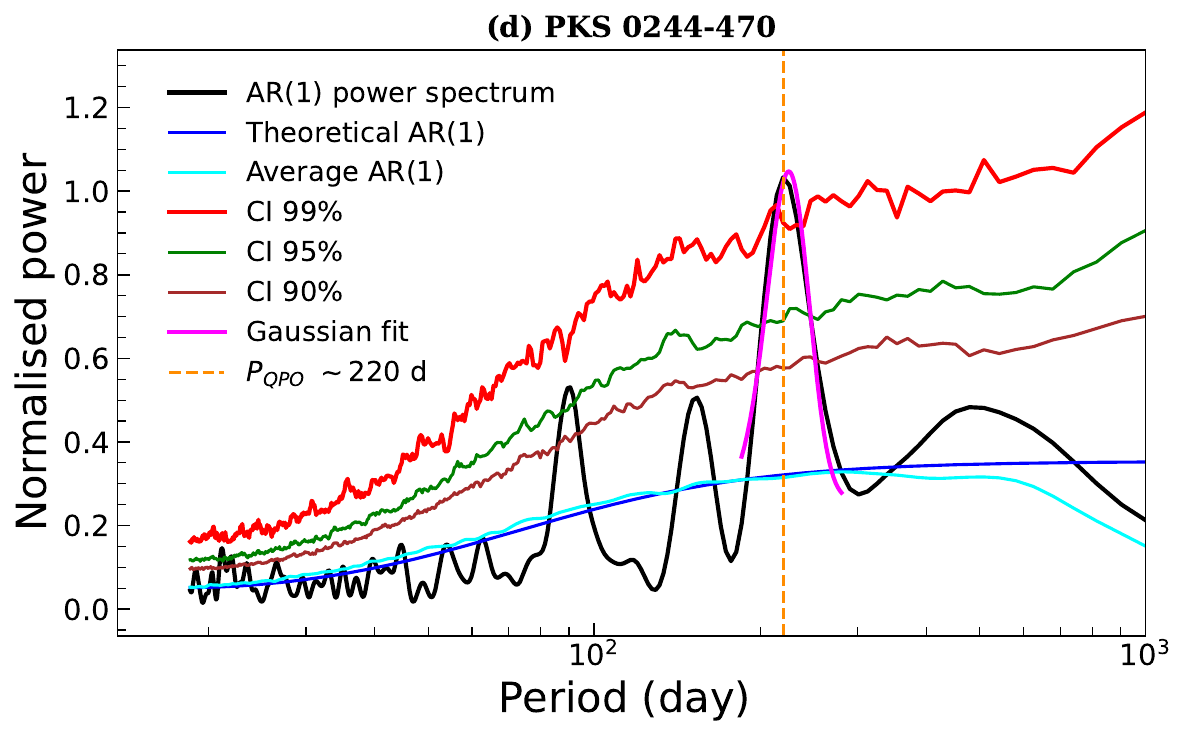}
    \vspace{1pt}
    \includegraphics[width=0.47\textwidth]{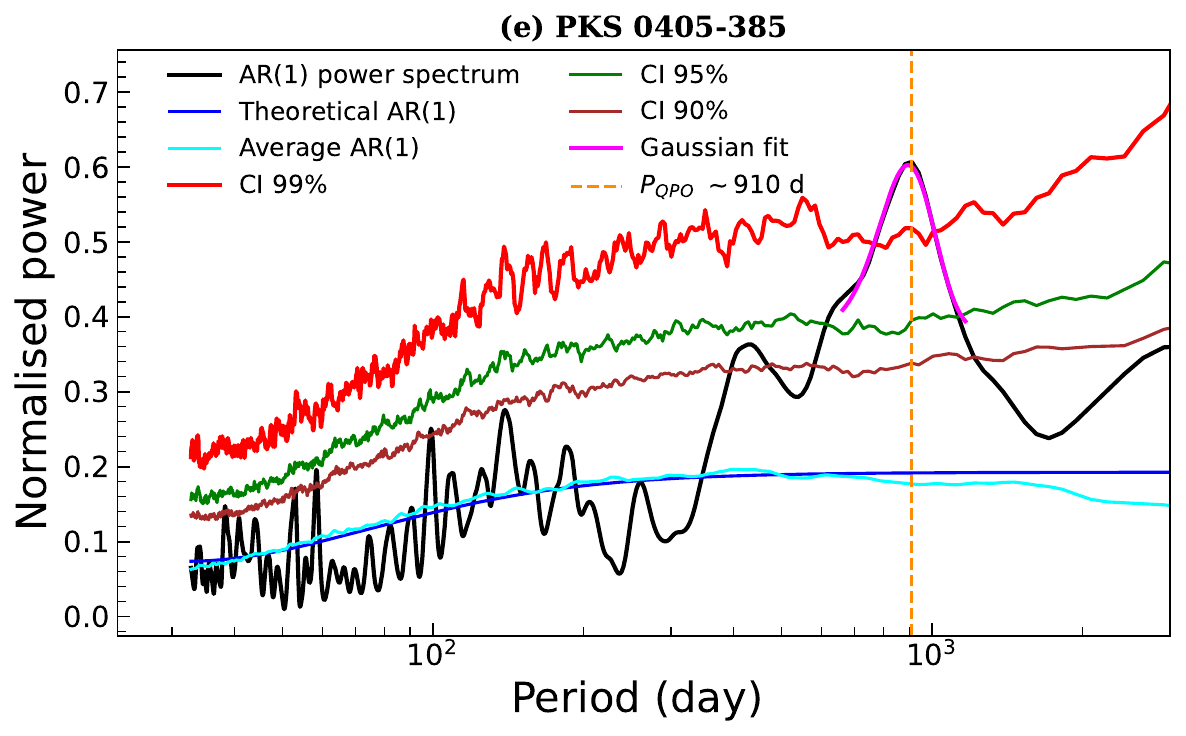} \hspace{1pt}
    \includegraphics[width=0.47\textwidth]{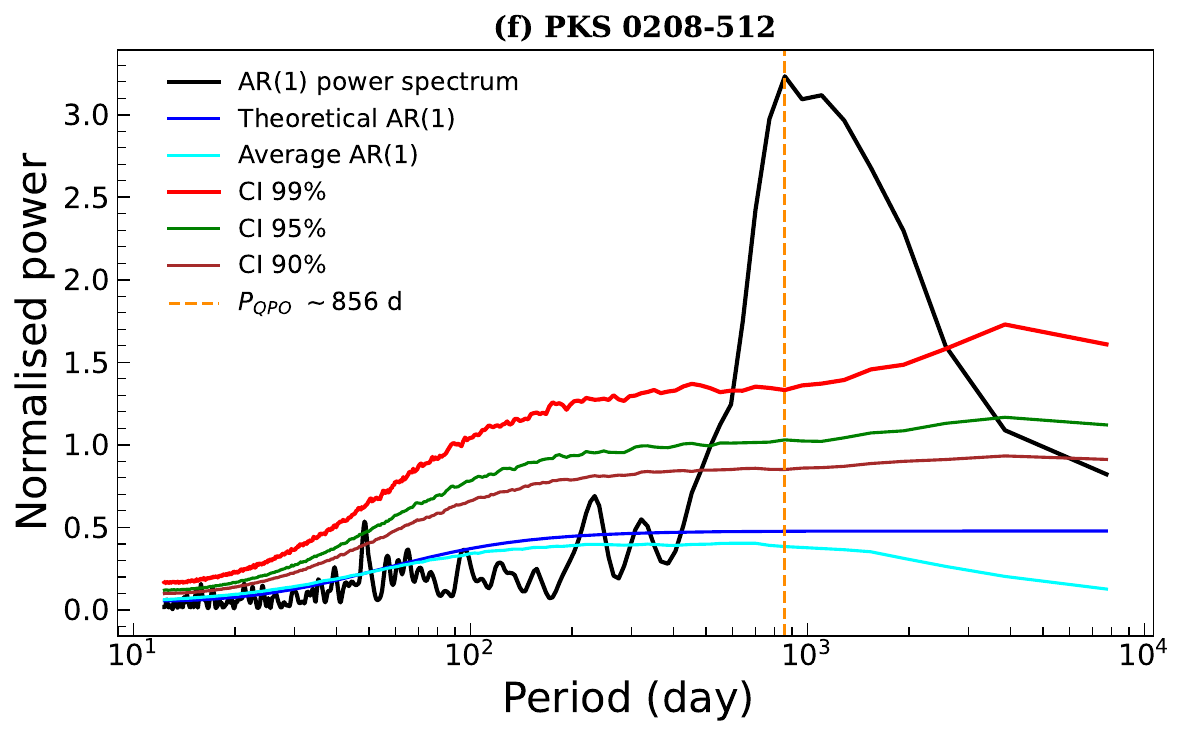}
    \vspace{1pt}
    \includegraphics[width=0.47\textwidth]{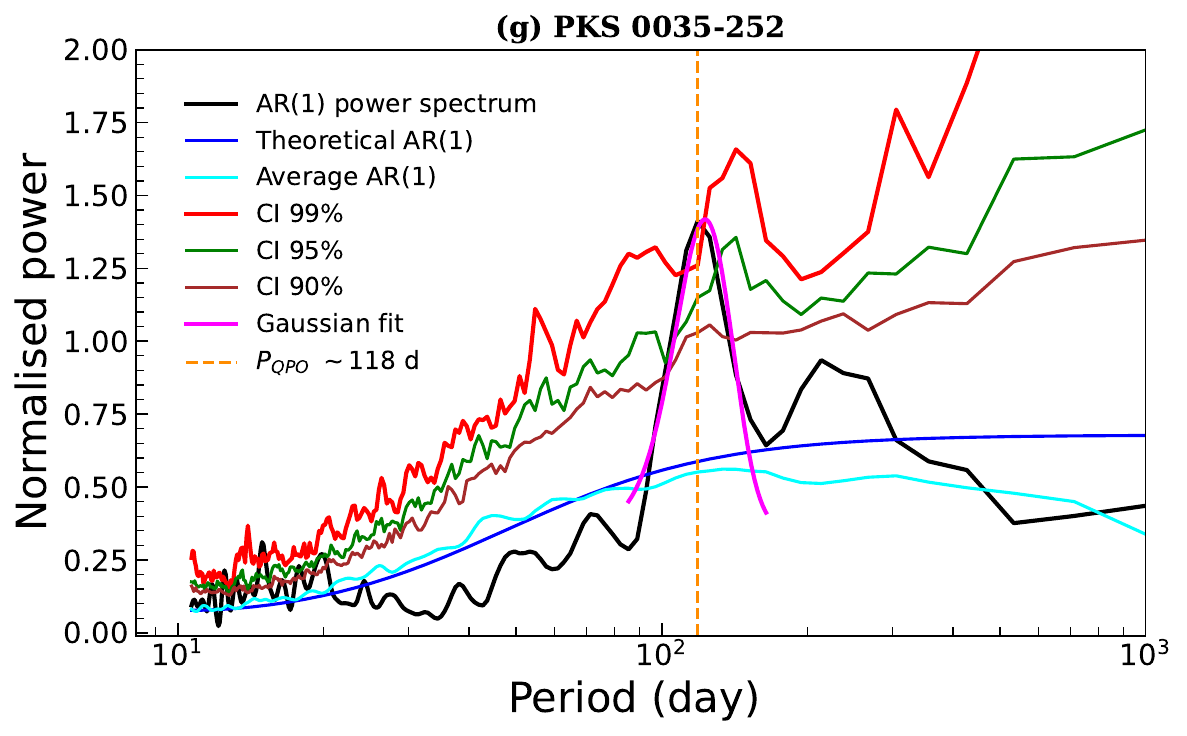}
    \caption{REDFIT analysis of $\gamma$-ray light curves of blazars, using the AR(1) process with the REDFIT software. The red noise-corrected power spectrums (black) are presented alongside theoretical (blue) and average red noise (cyan) spectrums. The significance levels of 99$\%$, 95$\%$, and 90$\%$ are indicated in red, green, and brown, respectively. Gaussian fits to the dominant peaks in magenta and peak positions are denoted by dotted vertical lines in orange.}
    \label{Fig-REDFIT}    
\end{figure*}

\begin{table*}
\setlength{\extrarowheight}{7pt}
\setlength{\tabcolsep}{8pt}
\centering
\begin{threeparttable}
\caption{Summary of the observed QPO findings of individual blazars. }
\label{tab:QPO}
\begin{tabular}{c c c c c c c }
\hline
\hline
Source  & Time span (MJD) & \multicolumn{2}{c}{LSP} & \multicolumn{2}{c}{REDFIT} & QPO cycle  \\
\cline{3-4} \cline{5-6}
 & & $\mathrm{f_{obs}} \ (\times \ 10^{-2} \ day^{-1})$ & Local significance & $\mathrm{P_{obs}} \ (\mathrm{yr})$ & Significance & \\
(1) & (2) & (3) & (4) & (5) & (6) & (7) \\
[+2pt]
\hline
PKS 0736+01 & 54721$ \ - \ $60502 &  0.069$\pm$0.0091 & $> 4\sigma$ & 4.4$\pm$1.08& $> 99\%$ & $\sim$4 \\ 
PKS 1424-41 & 56254$ \ - \ $58228 & 0.28$\pm$0.019 & $> 3\sigma$ & 1.0$\pm$0.13 & $> 99\%$ & 7 \\
S2 0109+22 & 57521$ \ - \ $60769 & 0.15$\pm$0.015 & $> 3\sigma$ & 1.78$\pm$0.45 & $> 99\%$ & 5  \\
PKS 0244-470 & 54693$ \ - \ $56317 & 0.44$\pm$0.035 & $> 3\sigma$ & 0.60$\pm$0.063 & $> 99\%$ & 7 \\
PKS 0405-385 & 54686$ \ - \ $60482 & 0.10$\pm$0.0074 & $\approx 3\sigma$ & 2.49$\pm$0.36 & $> 99\%$ & 6\\
PKS 0208-512 & 54714$ \ - \ $58563 & 0.11$\pm$0.010 & $\approx 3\sigma$ & 2.34$\pm \ -$ & $> 99\%$ & 5\\
PKS 0035-252 & 58386$ \ - \ $59454 &0.28$\pm$0.019 & $\approx 3\sigma$ & 0.32$\pm$0.05 & $> 99\%$ & 5  \\
[+5pt]
\hline
\end{tabular}
\begin{tablenotes} 
\small
\item Note: Summary of the quasi-periodic oscillation (QPO) results from the analysis of $\gamma$-ray light curves of the blazars in our sample. Column (1): Source name; Column (2): Duration over which QPO is detected; Column (3): QPO frequency identified from the LSP analysis; Column (4): Local significance of the LSP-detected QPO; Column (5): QPO period from REDFIT analysis; Column (6): Significance of the REDFIT-detected QPO; Column (7): Number of observed QPO cycles in the $\gamma$-ray light curve.

\end{tablenotes}
\end{threeparttable}
\end{table*}

The REDFIT analysis revealed prominent QPO features in the $\gamma$-ray light curves of blazars, with the corresponding oscillation period summarized as follows: PKS 0736+01 exhibits a QPO period of $4.4 \pm 1.08 \ \mathrm{yr}$, PKS 1424-41 shows $1.0 \pm 0.13 \ \mathrm{yr}$, S2 0109+22 has $1.78 \pm 0.45 \ \mathrm{yr}$, PKS 0244-470 displays $0.60 \pm 0.063 \ \mathrm{yr}$, PKS 0405-385 shows $2.49 \pm 0.36 \ \mathrm{yr}$, PKS 0208-512 has $2.34 \ \mathrm{yr}$, and PKS 0035-252 shows $0.32 \pm 0.05 \ \mathrm{yr}$. The uncertainties on the detected periods were estimated by fitting the QPO peaks with Gaussian functions and adopting the half-width at half-maximum (HWHM) as an error on the QPO period, except for PKS 0208-512. The findings are shown and tabulated in Figure~\ref{Fig-REDFIT} and Table~\ref{tab:QPO}, respectively.

\subsection{\rm{Gaussian process modelling}}\label{sec:Gaussian_process_DRW}
The AGN's variability is inherently stochastic in nature. The light curves can be well described by the stochastic processes, also known as Continuous Time Autoregressive Moving Average [CARMA(p, q)] processes \citep{kelly2014flexible}, defined as the solutions to the stochastic differential equation:

\begin{equation}
\begin{split}
\frac{d^p y(t)}{dt^p} + \alpha_{p-1}\frac{d^{p-1}y(t)}{dt^{p-1}}+...+\alpha_0 y(t) =\\
\beta_q \frac{d^q \epsilon(t)}{dt^q}+\beta_{q-1}\frac{d^{q-1}\epsilon(t)}{dt^{q-1}}+...+\beta_0 \epsilon(t),
\end{split}
\end{equation}

where, y(t) represents a time series, $\epsilon$(t) is a continuous time white noise process, and $\alpha_*$ and $\beta_*$ are the coefficients of autoregressive (AR) and moving average (MA) models, respectively. Here, p and q  are the order parameters of AR and MA models, respectively. \par
The simplest model is a continuous autoregressive [CAR(1)] model, also known as the Ornstein-Uhlenbeck process. It is a popular red noise model \citep{kelly2009variations, kozlowski2009quantifying, macleod2012description, ruan2012characterizing, zu2013quasar, moreno2019stochastic, burke2021characteristic, zhang2022characterizing, zhang2023gaussian, sharma2024probing, zhang2024discovering, sharma2024microquasars}, usually referred to as Damped Random Walk (DRW) model, described by the following differential equation:

\begin{equation}
    \left[ \frac{d}{dt} + \frac{1}{\tau_{DRW}} \right] y(t) = \sigma_{DRW} \epsilon(t)
\end{equation}

where $\tau_{DRW}$  and $\sigma_{DRW}$ are the characteristic damping time-scale and amplitude of the DRW process, respectively. The mathematical form of the covariance function of the DRW model is defined as

\begin{equation}
    k(t_{nm}) = a \cdot \exp(-t_{nm} \, c),
\end{equation}

where $t_{nm} = | t_n -t_m|$ denotes the time lag between measurements m and n, with $a = 2 \sigma_{DRW}^2$ and $c = \frac{1}{\tau_{DRW}}$. The power spectral density (PSD) of the DRW model is defined as:
\begin{equation}
    S(\omega) = \sqrt{\frac{2}{\pi}} \frac{a}{c} \frac{1}{1 + (\frac{\omega}{c})^2}
\end{equation}

The DRW PSD  has a form of Broken Power Law (BPL), where the broken frequency $f_b$ corresponds to the characteristic damping timescale $\tau_{DRW} = \frac{1}{2\pi f_b}$.\par
In the best-fit parameters estimation of the DRW model for both light curves, we employed the Markov chain Monte Carlo (MCMC) algorithm provided by the \textsc{emcee}\footnote{\url{https://emcee.readthedocs.io/en/stable/}} package \citep{foreman2013emcee}. For the modeling, we employed the \textsc{EzTao}\footnote{\url{https://eztao.readthedocs.io/en/latest/index.html}} package, which is built on top of \textsc{celerite}\footnote{\url{https://celerite.readthedocs.io/en/stable/}}. In this study, we generated the distributions of the posterior parameters by running 10000 steps as burn-in and 30000 as burn-out, which are shown in Figure~\ref{Fig-DRWCorner} and listed in Table~\ref{tab:DRW}.

\begin{figure*}
    \centering\includegraphics[width=0.52\textwidth]{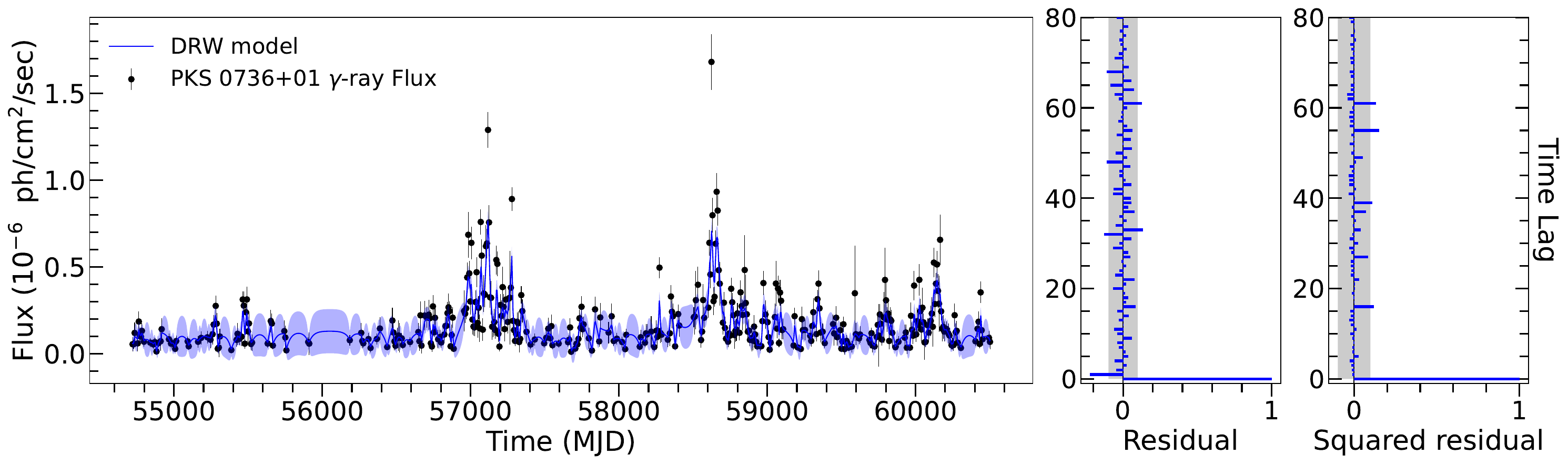} \vspace{1pt}
    \includegraphics[width=0.52\textwidth]{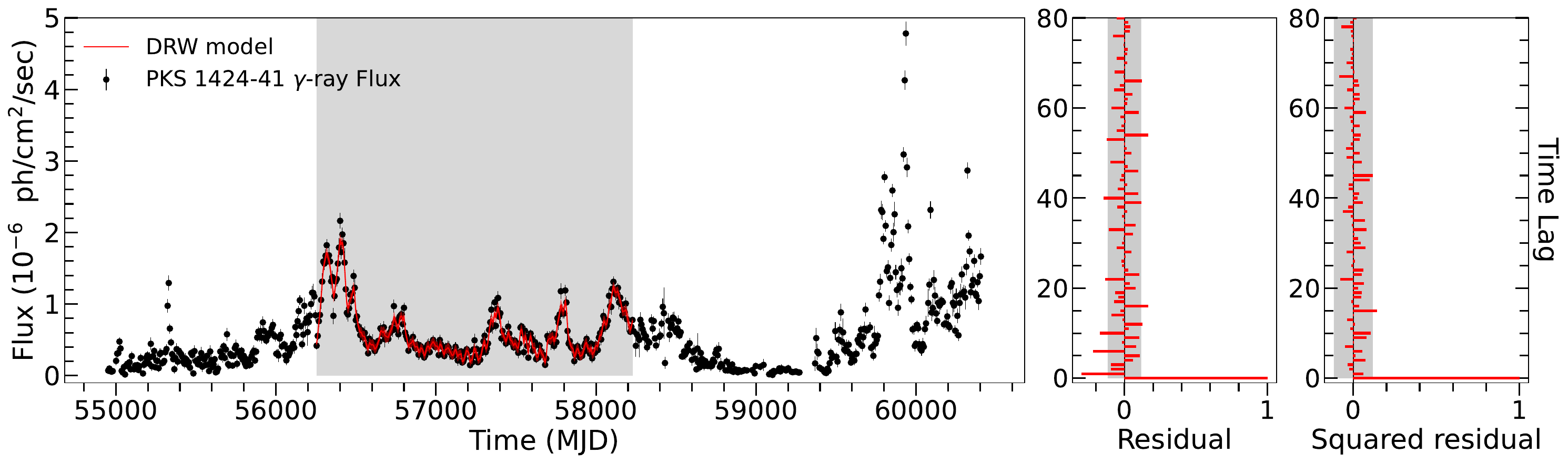}\vspace{1pt}
    \includegraphics[width=0.52\textwidth]{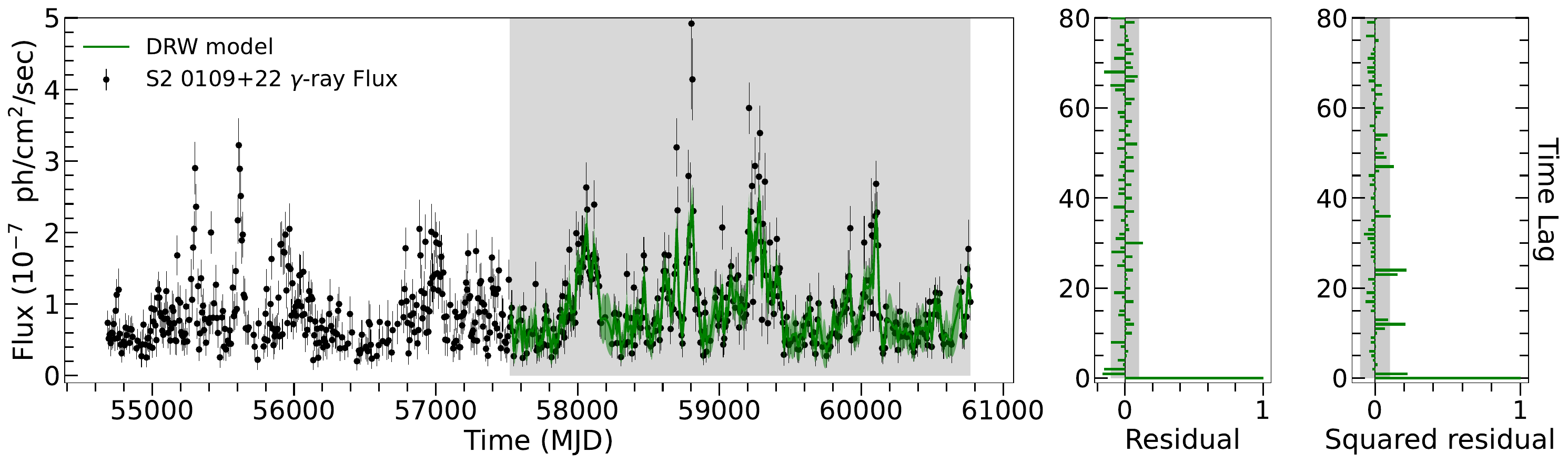}\vspace{1pt}
    \includegraphics[width=0.52\textwidth]{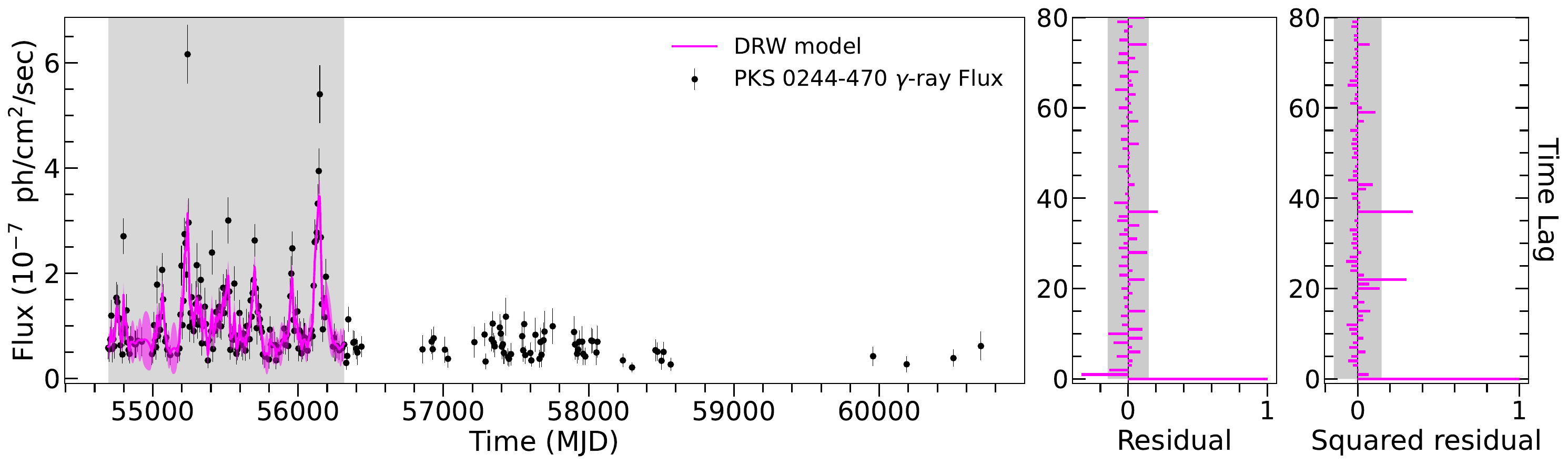}\vspace{1pt}
    \includegraphics[width=0.52\textwidth]{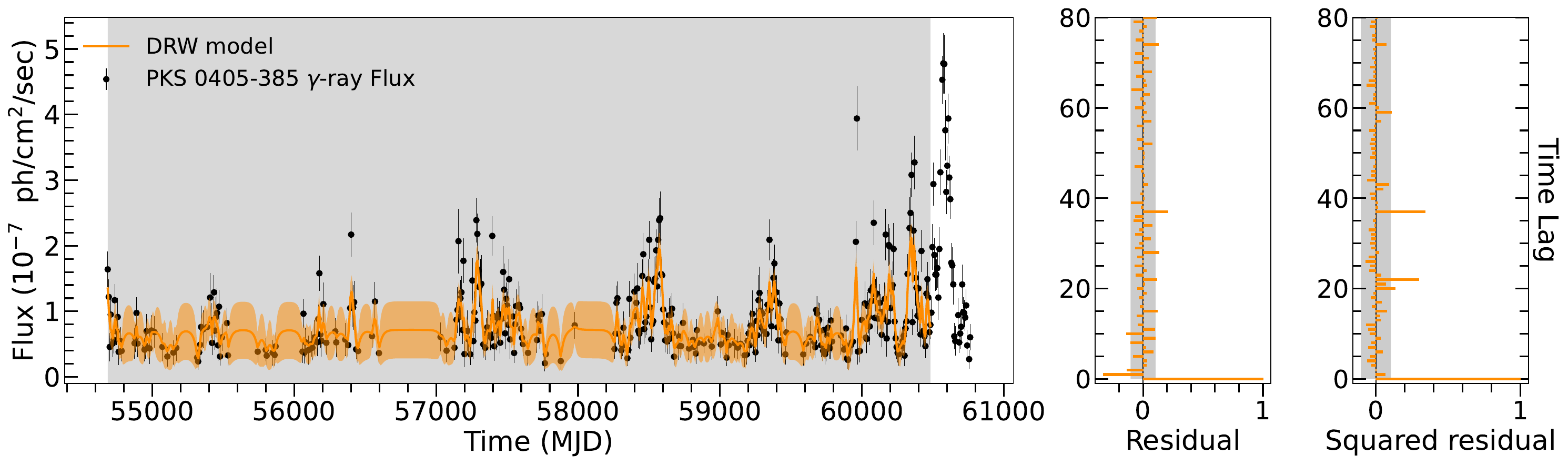} \vspace{1pt}
    \includegraphics[width=0.52\textwidth]{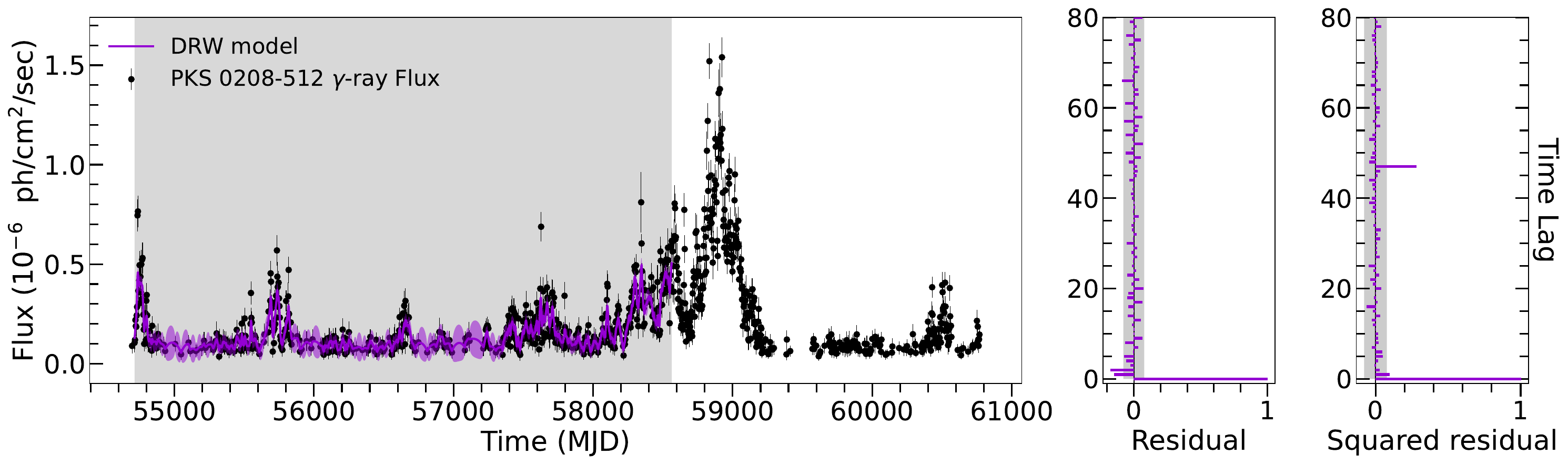}\vspace{1pt}
    \includegraphics[width=0.52\textwidth]{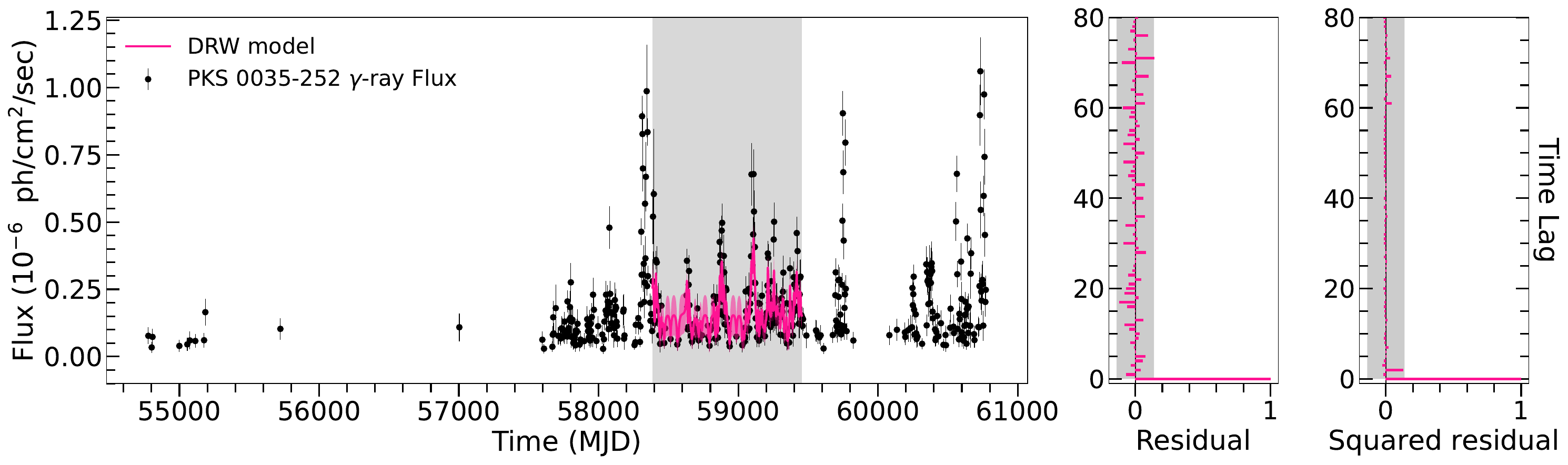}
    \caption{The DRW modeling was performed using the 7-day binned $\gamma$-ray light curve of blazars over the durations (grey shaded regions) listed in Table~\ref{tab:QPO}. The leftmost panel displays the $\gamma$-ray light curves, along with the best-fitting DRW model profile in different colors, including the 1$\sigma$ confidence interval. The middle and right panels show the autocorrelation functions (ACFs) of the standardized residuals (bottom left) and the squared of standardized residuals in their corresponding colors, respectively, along with 95$\%$ confidence intervals of the white noise. }
    \label{Fig-DRW1}    
\end{figure*}

\begin{figure*}
    \centering
    \includegraphics[width=0.47\textwidth]{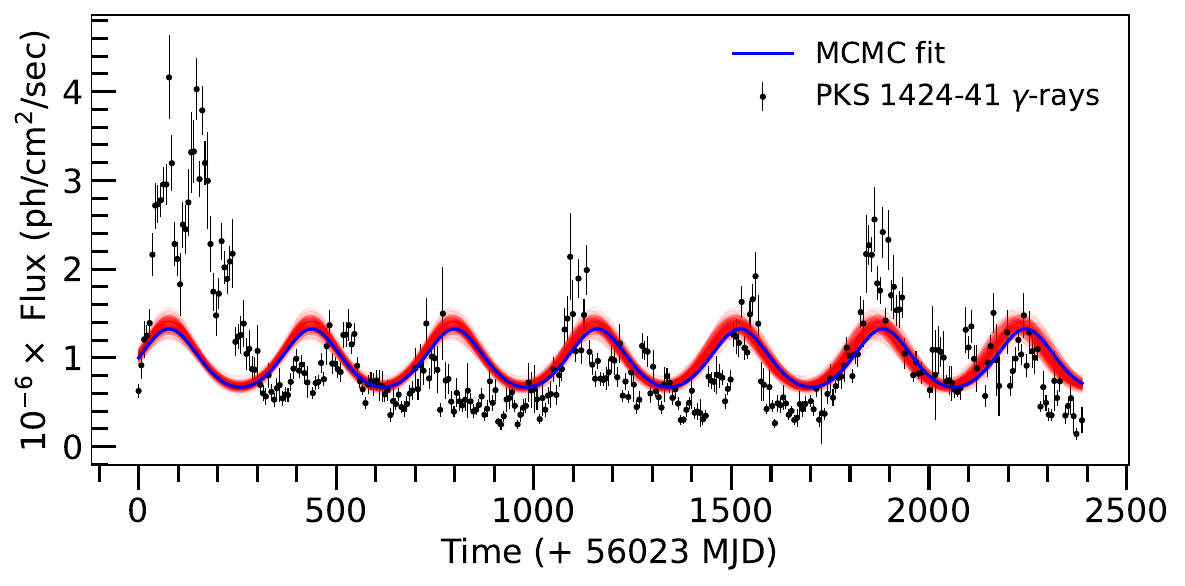} 
    \hspace{1pt}
    \includegraphics[width=0.47\textwidth]{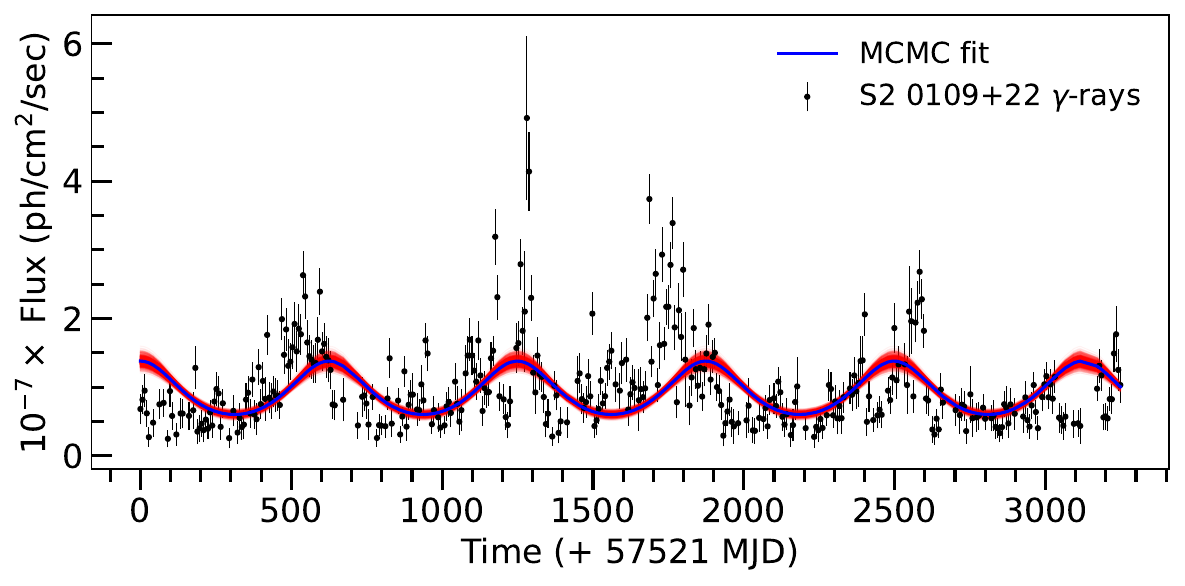}
    \vspace{1pt}
    \includegraphics[width=0.47\textwidth]{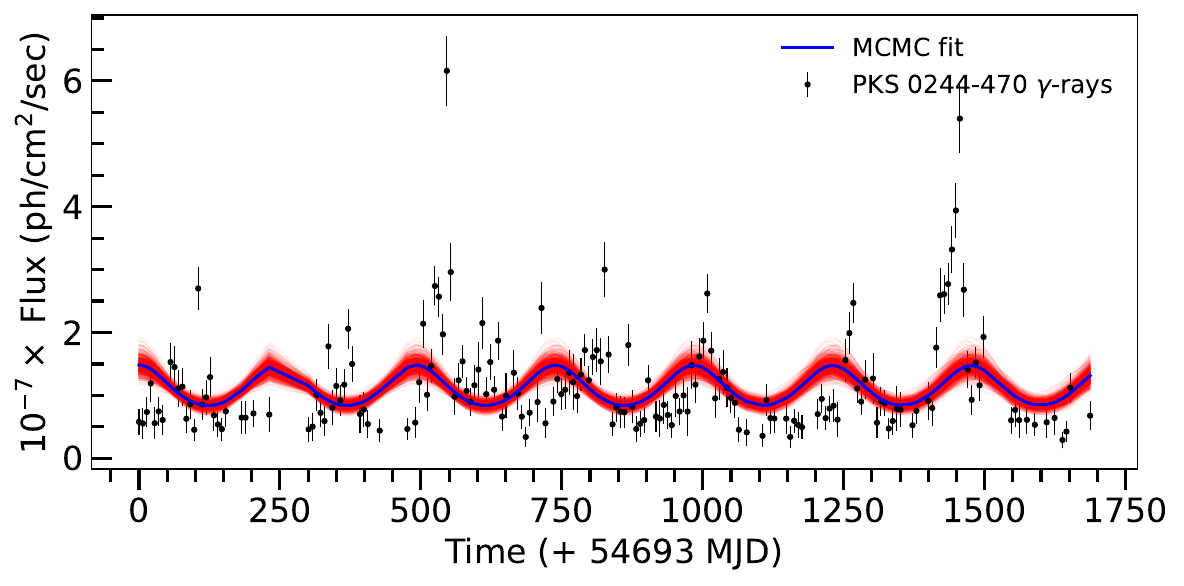}
    \hspace{1pt}
    \includegraphics[width=0.47\textwidth]{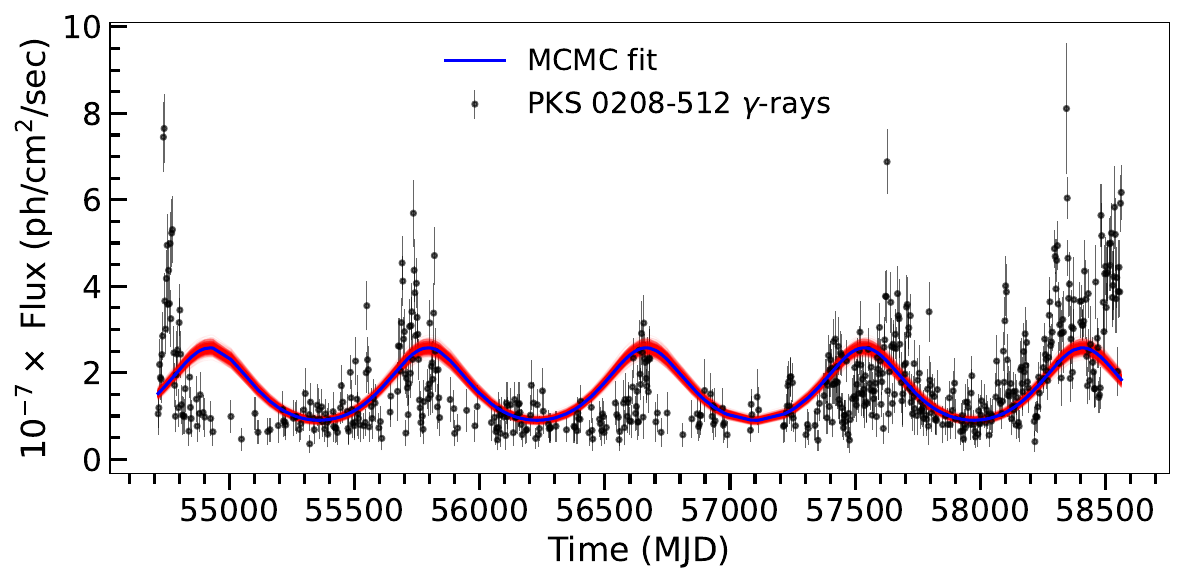}
    \vspace{1pt}
    \includegraphics[width=0.47\textwidth]{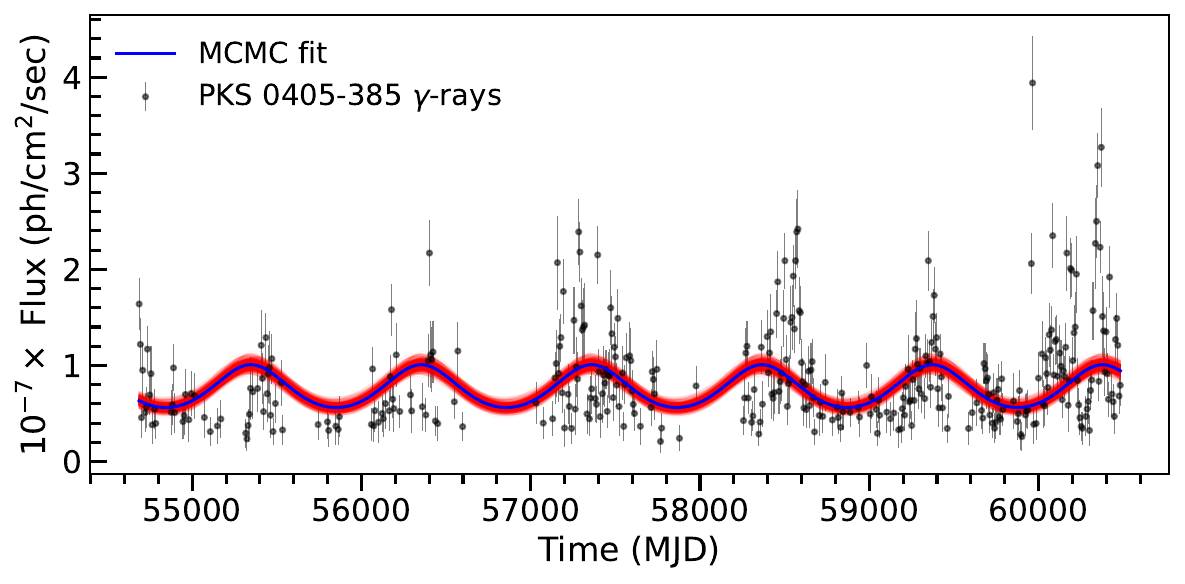}
    \caption{The $\gamma$-ray light curves displaying long-timescale QPOs are modeled under the scenario of a supermassive binary black hole (SMBBH) system, where one of the black holes launches a relativistic jet. The observed flux data are shown in black. The 1000 red curves are randomly selected from posterior samples, fitted to the $\gamma$-ray light curves of blazars through MCMC analysis. The blue curve indicates the mean model derived from all posterior samples, providing the best average representation of the QPO behavior.   }
    \label{Fig-mcmc_lc}    
\end{figure*}


\begin{figure*}
    \centering
    \includegraphics[width=0.4\textwidth]{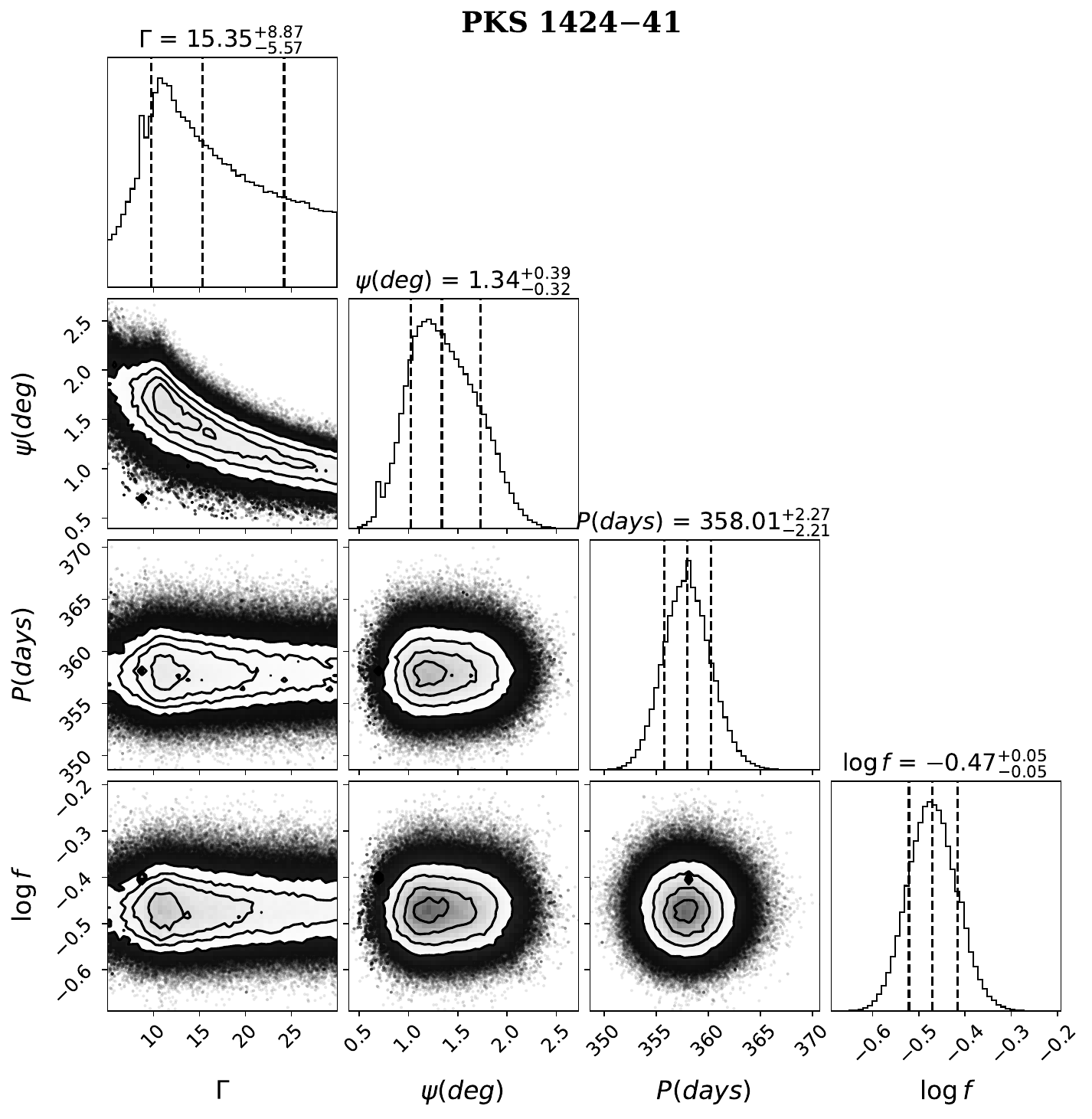} \hspace{1pt} 
    \includegraphics[width=0.4\textwidth]{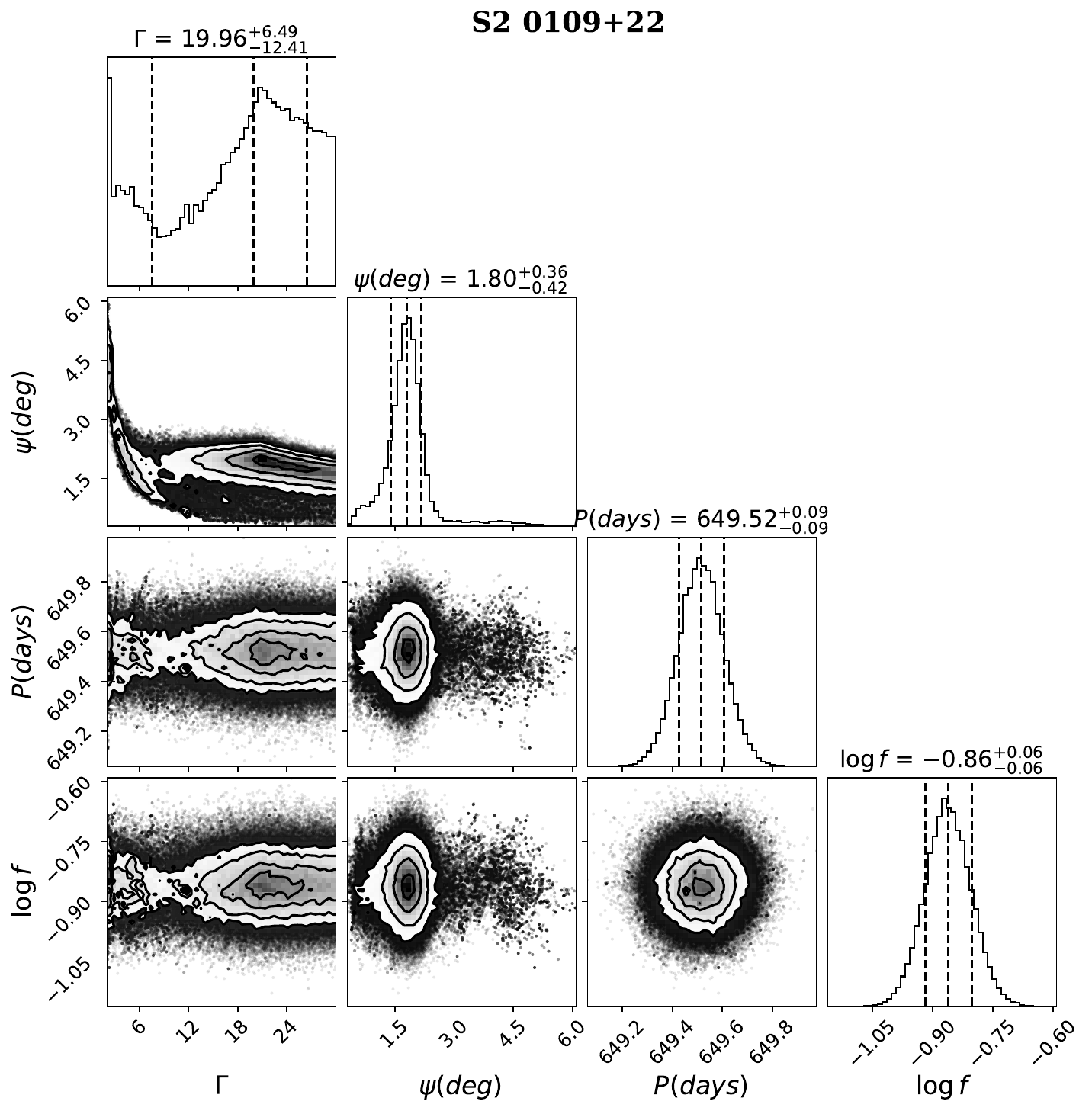}
    \vspace{1pt}
    \includegraphics[width=0.4\textwidth]{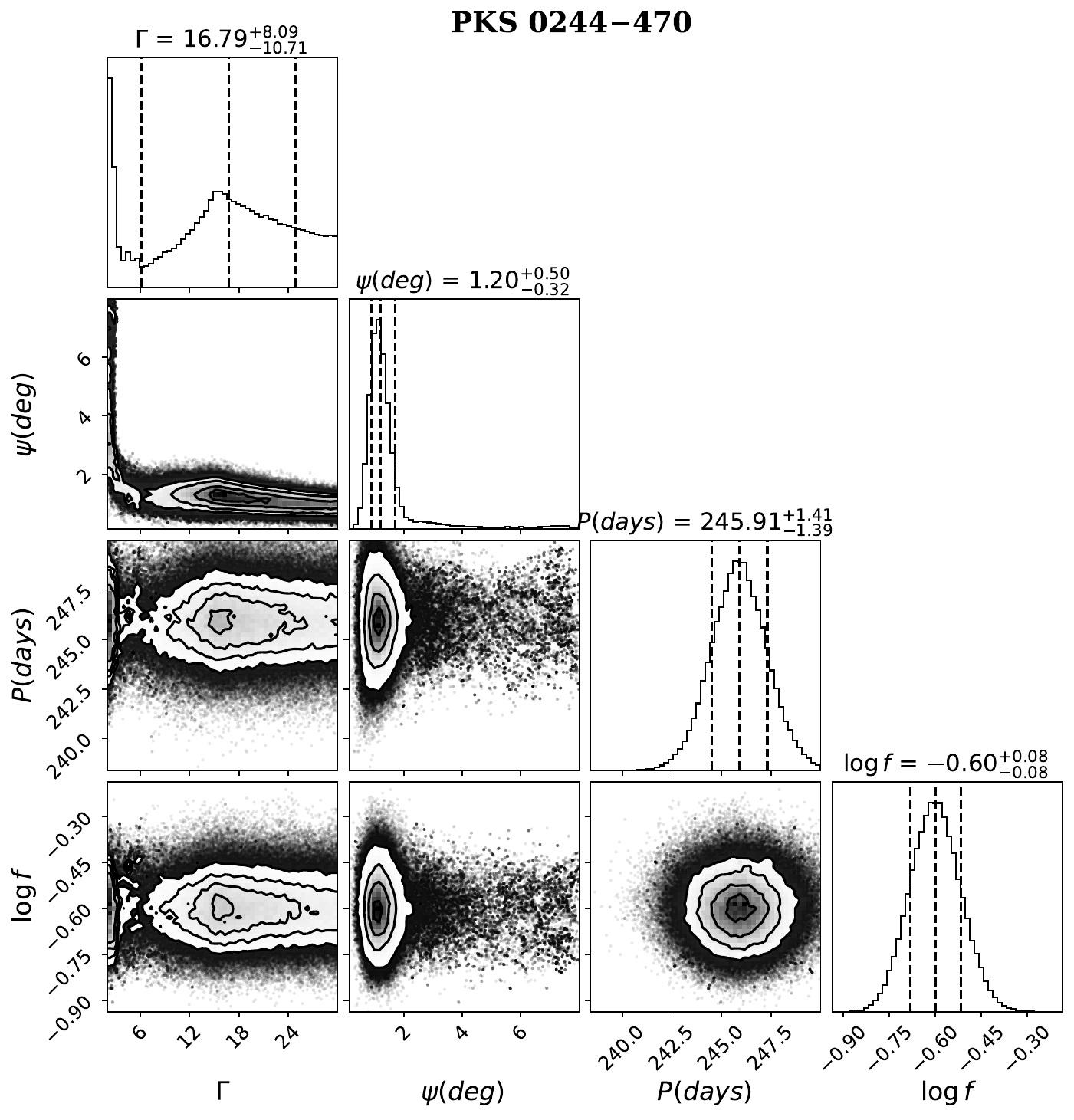}
    \hspace{1pt}
    \includegraphics[width=0.4\textwidth]{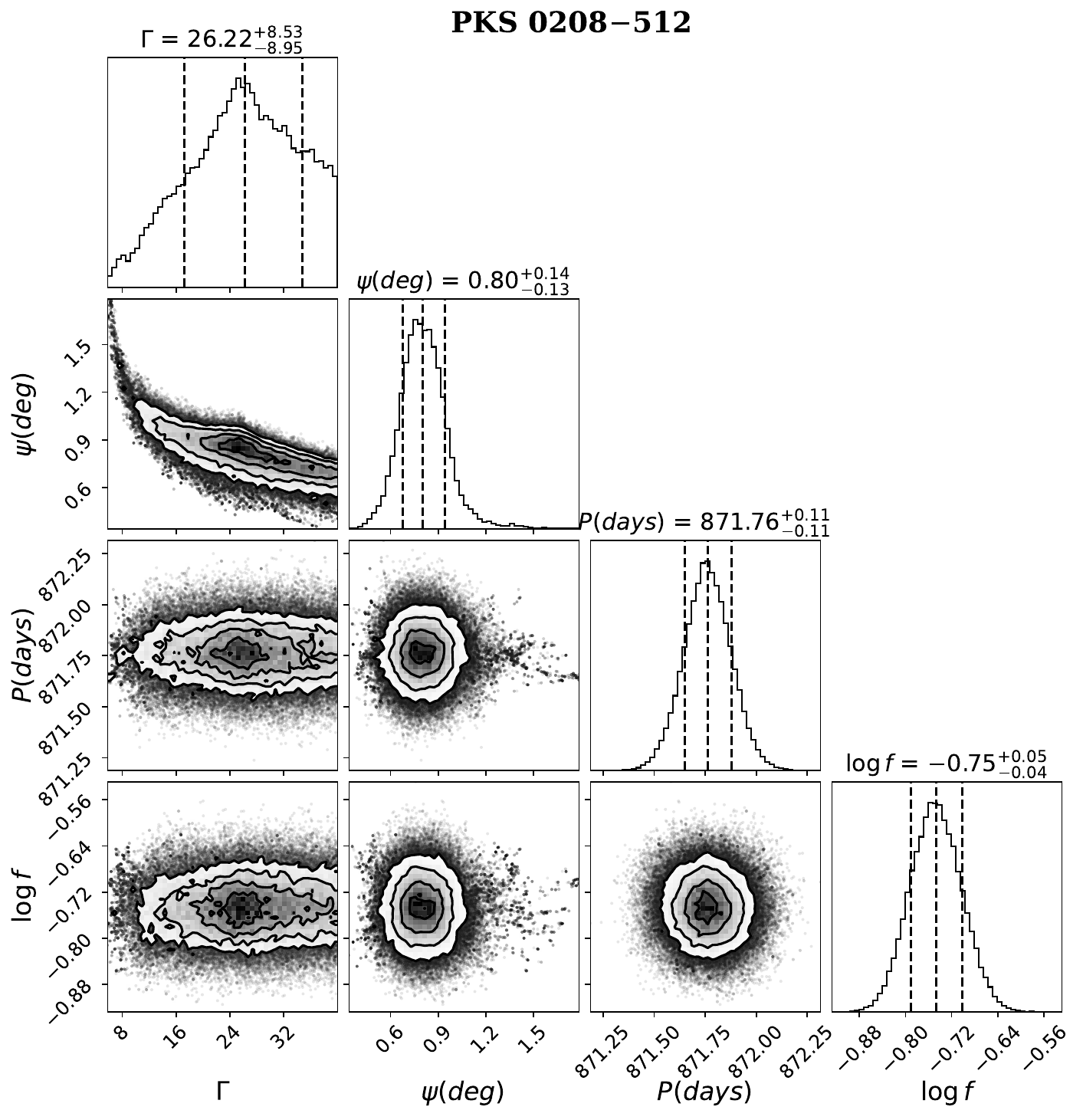}
    \vspace{1pt}
    \includegraphics[width=0.4\textwidth]{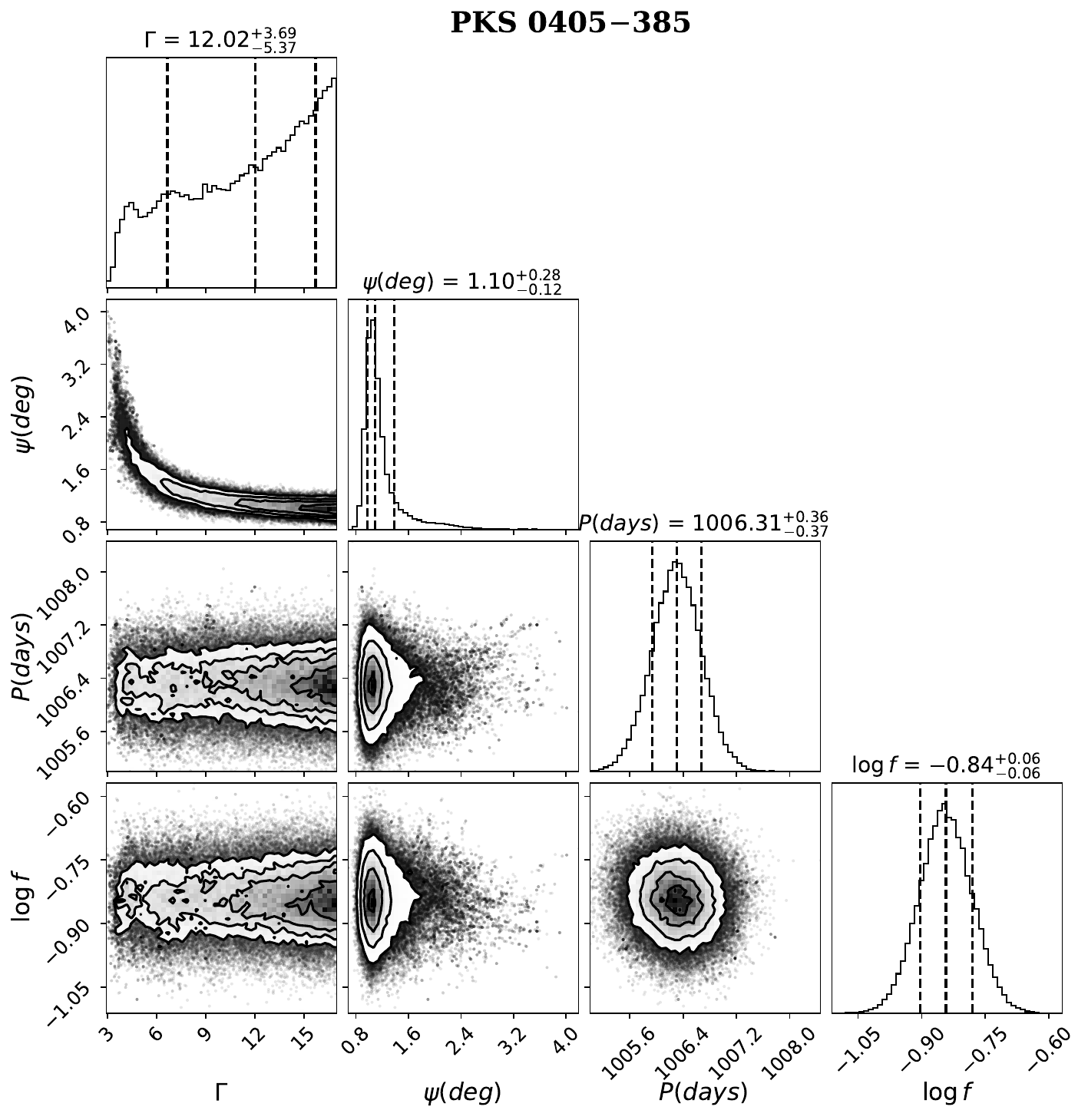}
    \caption{The posterior distribution of the model parameters derived under the supermassive binary black hole (SMBBH) framework.}
    \label{Fig-MCMC_LC_fitting}    
\end{figure*}


\subsection{\rm{Significance of QPO}}
AGN emissions are known to exhibit stochastic variability, often well-described by a red noise process. However, the presence of red noise, combined with uneven temporal sampling in the light curves, can produce artificial peaks in periodograms, potentially mimicking true periodic signals. Thus, accurately assessing the significance of any detected periodicity is essential.

To evaluate the significance of the observed periodic features in the LSPs, we modeled AGN variability with a red noise process using the Damped Random Walk (DRW) model, which captures the stochastic nature of AGN emissions through two key parameters: the variability amplitude and the characteristic damping timescale. These parameters were optimized for each light curve. Utilizing the EzTao package, we then generated 30,000 synthetic light curves with the same temporal sampling as for the original light curves. For each simulated light curve, we calculated the corresponding Lomb-Scargle periodogram (LSP), applying the same analysis pipeline used for the original light curve, allowing for a robust statistical comparison. 

To estimate the significance level of the dominant peak in original $\gamma$-ray LSPs of all blazars in our sample, we calculated the 84th, 97.5th, 99.85th, and 99.995th percentiles of the 30000 mock LSPs for each candidate frequency value, which correspond to the 1$\sigma$, 2$\sigma$, 3$\sigma$, and 4$\sigma$ significance level. Our analysis shows that the dominant peak in the LSP of PKS 0736+01 exceeds the 4$\sigma$ significance level. The LSP peaks of the remaining sources- PKS 1424-41, S2 0109+22, and PKS 0244-470- surpass the 3$\sigma$ threshold, while PKS 0405-385, PKS 0208-512, and PKS 0035-252 exhibit significance levels of 2.9$\sigma$, 2.85$\sigma$, and 2.8$\sigma$, respectively. These results are consistent with the REDFIT analysis, where all detected quasi-periodic oscillation (QPO) peaks exceed the 99$\%$ confidence level. To further assess the reliability of the detected signals, we examined the spectral window periodograms. This was done by creating a light curve/time series over the same time span, but with a temporal resolution ten times finer. Time stamps corresponding to actual observations were set to one, and all others to zero. The LSP of this spectral window function (shown in pink in Figure~\ref{Fig-DRW_LSP}) helps identify spurious peaks that may arise from irregular sampling patterns. In our case, the $\gamma$-ray light curves are well-sampled, and the spectral window periodograms do not reveal any spurious signals, supporting the robustness of our detections. However, it is well known that periodogram analyses can be influenced by uneven sampling and finite-duration observations. Previous studies \citep{kozlowski2017limitations, suberlak2021improving} have highlighted how such factors can introduce biases into variability measurements. \citet{burke2021characteristic} specifically emphasized that a credible variability timescale must exceed the mean cadence and be less than 10$\%$ of the total baseline of the light curve. We adopt this criterion to identify unreliable regions in the LSPs, which are shown as grey-shaded areas in Figure~\ref{Fig-DRW_LSP}. These regions are excluded from the interpretation of significant QPO detections.

\section{\rm{Modeling of Gamma-ray light curves}}\label{sec:LC_modeling}
In this work, we have investigated the flux modulations in the $\gamma$-rays of blazars in our sample over the Fermi satellite operation and identified long-term and transient quasi-periodic behaviour ranging from a few months to years. The corresponding sections and the Table mention a detailed investigation and findings. Several physical models have been proposed in the literature to explain the phenomenon of quasi-periodic oscillations (QPOs) in blazars. These include three primary explanations for the QPOs involving the presence of a super-massive binary black hole system (SMBBH), jet precession/rotation, or helical motion of relativistic jet, and pulsation accretion flow instabilities. A more detailed discussion on each of these scenarios is given below. 

\begin{figure}
    \centering
    \includegraphics[width=0.5\textwidth]{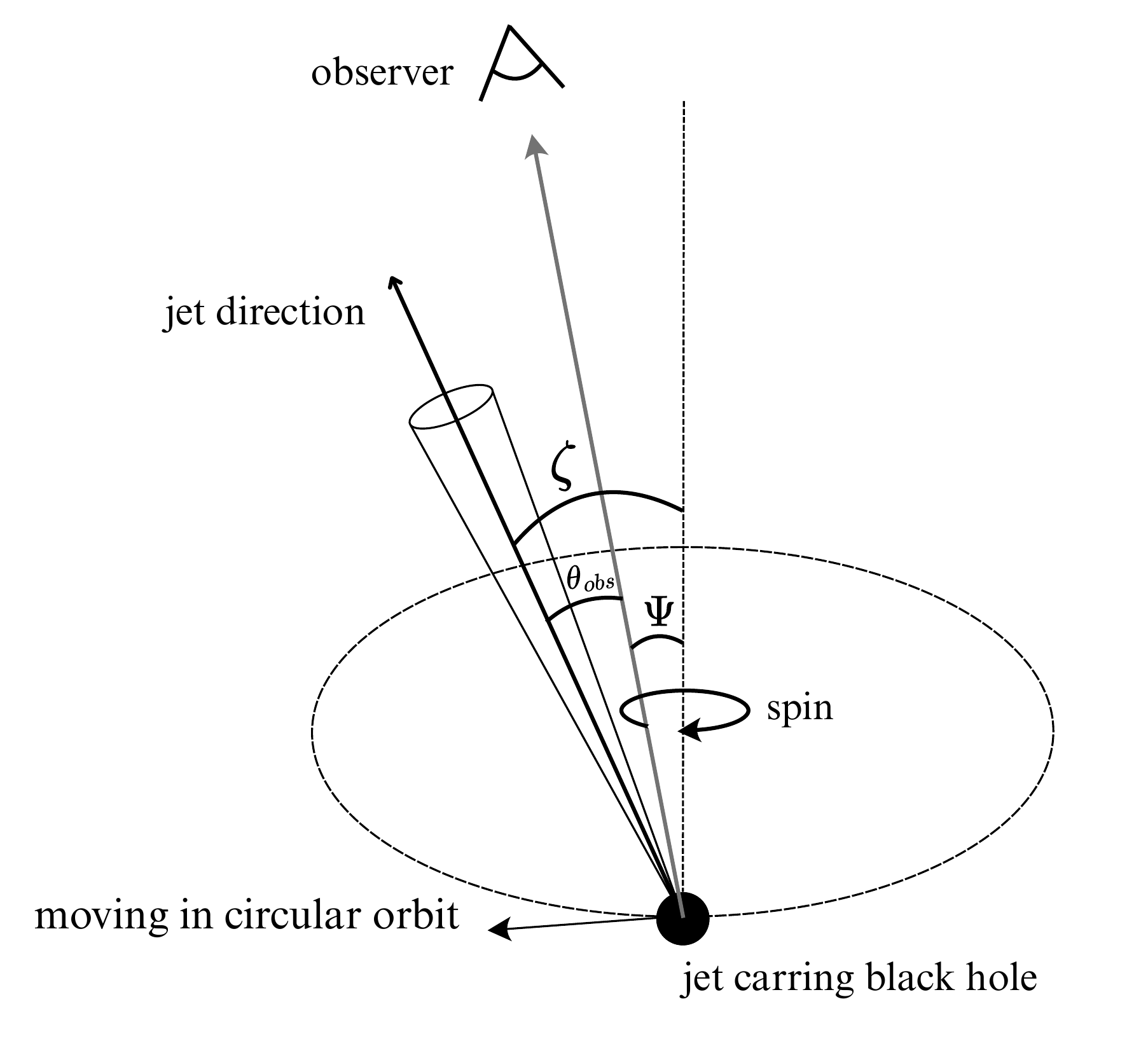} 
    \caption{Schematic representation of the SMBH and its associated jet. The long-dashed circular line illustrates the orbital path of the black hole within a supermassive binary system. The black hole's spin axis is assumed to be perpendicular to the orbital plane. In the center of mass frame of the binary, the jet is launched at an angle $\zeta$ with respect to the orbital angular momentum. The observer's line of sight forms an angle $\theta_{obs}$ with the jet axis, while $\psi$ denotes the angle between the line of sight and the black hole’s spin axis.}
    \label{Fig-Scematic_SMBBH}    
\end{figure}

\subsubsection{Jet scenario}\label{sec:jet_model}
Blazar emission is typically dominated by Doppler-boosted synchrotron radiation from relativistic jets, and variability is often attributed to disturbances propagating along these jets. This makes it plausible that the quasi-periodic oscillations (QPOs) observed in blazars are linked to jet-related processes. One compelling geometric interpretation involves a helical jet structure. In this model, relativistic beaming enhances the emission as a plasma blob follows a helical trajectory within the jet. The changing orientation of the blob relative to the observer leads to periodic variations in the viewing angle over time \citep{camenzind1992lighthouse, villata1999helical, rieger2004geometrical, Li_2009, Li_2015, mohan2015kinematics, 2017MNRAS.465..161S, zhou201834, otero2020quasi, gong2022quasiperiodic, gong2023two, prince2023quasi, sharma2024detection}. Such helical motion may arise from jet bending, precession, or intrinsic helical magnetic fields, all of which can induce periodic modulation in the observed flux.

The time-dependent viewing angle is given by:
\begin{equation}
\cos \theta_{\mathrm{obs}}(t) = \cos i \cos \alpha  \ + \ \sin i \sin \alpha \cos \left( \frac{2\pi t}{P_{\mathrm{obs}}} \right) ,
\end{equation}
where $\alpha$ is the pitch angle between the blob's velocity and the jet axis, $i$ is the angle between the jet axis and the observer’s line of sight, and $P_{\mathrm{obs}}$ is the observed period of the oscillation.

Due to the helical motion, the Doppler factor varies with time as:
\begin{equation}
\delta(t) = \frac{1}{\Gamma \left(1 - \beta \cos \theta_{\mathrm{obs}}(t)\right)},
\end{equation}
where $\Gamma = 1/\sqrt{1 - \beta^2}$ is the bulk Lorentz factor and $\beta = v_{\mathrm{jet}}/c$ is the normalized jet speed. The observed flux is then modulated as $F_{\nu} \propto \delta(t)^3 F'_{\nu'}$.

The observed period is related to the rest-frame period of the blob motion via:
\begin{equation}
P_{\mathrm{rest}} = \frac{P_{\mathrm{obs}}}{1 - \beta \cos i \cos \alpha}.
\end{equation}

However, in single supermassive black hole jet scenarios, the expected range of QPO timescales is typically $P_{\mathrm{obs}}\sim 1$–130 days \citep{rieger2004geometrical, mohan2015kinematics}. In contrast, the $\gamma$-ray QPOs identified in this study generally exceed 130 days (except for one source), suggesting that the single-jet helical motion scenario may be insufficient to account for such long-timescale periodicities. In this study, month-like QPO timescales have been found in PKS 0035-252 and PKS 0244-470 with timescales of $\sim 3.7$ months and $\sim 7.5$ months, respectively, which suggests that the moving magnetized plasma blob in helical trajectory can cause the QPO behavior in the $\gamma$-ray emissions.
\begin{itemize}
    \item The $\gamma$-ray light curve of PKS 0035-252 reveals a transient-like quasi-periodic oscillation (QPO) spanning from MJD 58386 to 59454, with an observed timescale of approximately $\sim112$ days. To estimate the intrinsic period in the rest frame $(P_{rest})$, we adopted typical values of $\alpha=2^{\circ}, \ \ i=5^{\circ}$, and $ \Gamma = 10$ \citep{zhou201834, sarkar2019long, prince2023quasi, sharma2024detection, sharma2025searching}. This yields a rest-frame period of $P_{rest}$ $\sim$32.334 yr. The corresponding distance traveled by a relativistic blob during one cycle is estimated as $D = c\beta \ P_{rest} \ \mathrm{cos(\phi)} \approx 9.85 \ pc$ and the total projected distance over five cycles is $D_{p} = 5D \mathrm{sin(\psi)}\approx 4.29 \ pc$.

    \item Similarly, PKS 0244-470 exhibits a QPO in its $\gamma$-ray light curve between MJD 54693 and 56317, with an observed oscillation timescale of about  $\sim227$ days. Using the same set of parameters ( $\alpha=2^{\circ}, \ \ i=5^{\circ}$, and $ \Gamma = 10$), we find the intrinsic period to be $P_{rest}$ $\sim$65.53 yr, corresponding to a distance traveled per cycle of $D$$\approx19.97 \ pc$, and a projected seven-cycle distance of $D_p$$\approx12.18 \ pc$. As mentioned earlier, the single-jet helical motion scenario may not adequately account for QPOs with timescales exceeding a few months. To address this, we modeled the $\gamma$-ray light curve within the framework of a supermassive binary black hole (SMBBH) system, wherein one of the black holes is assumed to launch a relativistic jet. A comprehensive discussion of this model and the associated findings is provided in Section~\ref{sec:SMBBH}. 
\end{itemize}

\subsubsection{Disc scenario}\label{sec:disc_scenario}
The observed QPOs may also originate from instabilities or fluctuations within the accretion disk, which are efficiently transmitted to the jet and modulate its non-thermal emission. In this scenario, processes such as oscillations in the innermost regions of the accretion disk or Kelvin-Helmholtz instabilities can lead to quasi-periodic injections of plasma into the jet, thereby inducing periodic variations in the observed jet emission \citep{gupta2008periodic, 10.1093/mnras/stu1135, bhatta2016detection, sandrinelli2016quasi, tavani2018blazar}.

The mass of the central supermassive black hole (SMBH) can be estimated under this model using the following relation:
\begin{equation}
M = \frac{3.23 \times 10^4 \ \ P_{\mathrm{obs}}}{(r^{3/2} + a)(1 + z)} \ M_{\odot},
\label{eq:disc}
\end{equation}
where $P_{\mathrm{obs}}$ is the observed QPO period in seconds, $\delta$ is the Doppler factor, $r$ is the radius of the emission region in units of $GM/c^2$, $a$ is the dimensionless spin parameter of the SMBH, and $z$ is the redshift of the source. Following equation~\ref{eq:disc}, we estimated masses of SMBH of all blazars for a Schwarzschild ($M_{Sch}$) and Kerr ($M_{Kerr}$) black hole by adopting parameters, r=6 and a=0 for $M_{Sch}$, and r=1.2 and a=0.9982 for $M_{Kerr}$ for all blazars.
\begin{itemize}
    \item PKS 0035-252: The estimated SMBH masses are approximately $\sim 1.42\times 10^{10} \ \mathrm{M_{\odot}}$ for the Schwarzschild black hole and $\sim 9.07\times 10^{10} \ \mathrm{M_{\odot}}$ for a Kerr black hole.
    
    \item PKS 0736+01: In this case, the estimated black hole masses, $\mathrm{M_{Sch}}$ and $\mathrm{M_{Kerr}}$ are $\sim 2.31\times 10^{11} \ \mathrm{M_{\odot}}$ and $\sim 1.47\times 10^{12} \ \mathrm{M_{\odot}}$, respectively.
    
    \item PKS 1424-41: The estimated black hole masses, $\mathrm{M_{Sch}}$ and $\mathrm{M_{Kerr}}$ are $\sim 2.68\times 10^{10} \ \mathrm{M_{\odot}}$ and $\sim 1.7\times 10^{11} \ \mathrm{M_{\odot}}$, respectively.

    \item S2 0109+22: The estimated black hole masses are $\mathrm{M_{Sch}}$$\sim 9.29\times 10^{10} \ \mathrm{M_{\odot}}$ and $\mathrm{M_{Kerr}}$$\sim 5.90\times 10^{11} \ \mathrm{M_{\odot}}$.

    \item PKS 0244-470: The estimated black hole masses are $\mathrm{M_{Sch}}$$\sim 1.80\times 10^{10} \ \mathrm{M_{\odot}}$ and $\mathrm{M_{Kerr}}$$\sim 1.14\times 10^{11} \ \mathrm{M_{\odot}}$.

    \item PKS 0405-385: The estimated black hole masses are $\mathrm{M_{Sch}}$$\sim 7.94\times 10^{10} \ \mathrm{M_{\odot}}$ and $\mathrm{M_{Kerr}}$$\sim 5.04\times 10^{11} \ \mathrm{M_{\odot}}$.

    \item PKS 0208-512: The estimated black hole masses are $\mathrm{M_{Sch}}$$\sim 9.61\times 10^{10} \ \mathrm{M_{\odot}}$ and $\mathrm{M_{Kerr}}$$\sim 5.47\times 10^{11} \ \mathrm{M_{\odot}}$.
\end{itemize}

The estimated black hole masses of all blazars for both cases, including the Schwarzschild and Kerr black holes, are substantially higher than the SMBH mass reported by \citet{pei2022estimation, fan2004black, zhang2023detection, shaw2012spectroscopy, gong2022quasiperiodic, ghisellini2010general}, indicating that this scenario may not plausibly account for the observed QPO behavior in the $\gamma$-ray emissions of blazars in our sample. 


\subsubsection{SMBBH scenario}\label{sec:SMBBH}
An SMBBH scenario is widely used to explain the long-term QPO behavior, which might be caused by the periodic accretion perturbations triggered by the Keplerian orbital motion of the SMBBH or by gravitational torque from the companion, which can cause the jet-precession in misaligned disk orbits \citep{ackermann2015multiwavelength, cavaliere2017blazar, li2023radio}. In the Keplerian orbital motion scenario, the orbital parameters of the binary system can be estimated, e.g., the semimajor axis a is given by $P_{int}^2 = \frac{4\pi^2 a^3}{G(m+M)}$, where G is the gravitational constant, $P_{int}$ is the intrinsic orbital period, which is redshift corrected as $P_{int} = \frac{P_{obs}}{1+z}$, M and m are the primary and secondary BH masses, respectively. \citep{rieger2004geometrical, rieger2007supermassive, steinle2024probing, sharma2025searching} investigated the potential geometrical origins of periodicity in the blazars. In particular, Newtonian-driven jet precession in a close SMBBH system may be well associated with the observed QPO with timescale $\ge 1$ yr. In addition, the jet precession can also arise from the Lense-Thirring disc precession of the inner edge of the disc \citep{fragile2009general}. However, the time scale of the jet precession due to the Lense-Thirring effect occurs characteristically over a long time scale ($\sim \mathrm{Myr}$) and hence this scenario may be irrelevant for detected QPOs. \par
The detected QPOs with long time-scales are probably associated with jet precession or non-ballistic helical motion driven by the orbital dynamics of a close SMBBH system. To explore the origin of the QPOs, we adopt a model based on the supermassive binary black hole (SMBBH) scenario, in which one of the black holes launches a relativistic jet \citep{sobacchi2016model}. This model is used to replicate the $\gamma$-ray light curves exhibiting long timescale QPOs and is schematically illustrated in Figure~\ref{Fig-Scematic_SMBBH}. Within this model, the time-varying cosine of angle, represented by $\mathrm{cos[\theta(t)]}$, is expressed as:

\begin{equation}
\cos [\theta_{\mathrm{obs}}(t)] = \sin \psi \sin \zeta \cos \left[ \frac{2\pi (t - t_0)}{P_{\mathrm{obs}}} \right] \ + \ \cos \psi \cos \zeta ,
\end{equation}

 where $\psi$ is the angle between the observer’s line of sight and the orbital angular momentum, $\zeta$ is the angle between the jet axis and the orbital angular momentum, and $t_0$ is an initial condition of time. To replicate the $\gamma$-ray emissions using this model, we employed the Markov-chain Monte Carlo (MCMC) approach. We maximize the likelihood function $\mathcal{L}$ corresponding to this model using the form 
\begin{equation}
    ln\mathcal{L} = -\frac{1}{2} \ \sum_{i=1}^n \ \left( \frac{ \left( y_i - \mu_i\right)^2}{s_i^2} + ln( 2\pi s_i^2) \right)
\end{equation}

where 
\begin{equation}
    s_i^2 = \sigma_i^2 + f^2(\mu_i^2)
\end{equation}
where $y_i$ are the observed fluxes, $\mu_i$ are the model-predicted fluxes based on the input parameters, and $\sigma_i$ denotes the measurement uncertainties. The parameter $f$ accounts for an additional fractional uncertainty in the flux estimates, representing underestimated observational errors. The key model parameters include the Lorentz factor ($\Gamma$), the angle between the line of sight and the spin axis ($\psi$), the angle between the jet axis and the spin axis ($\zeta$, fixed at $5^{\circ}$), the QPO period ($P$), the baseline flux ($F_0$), and the fractional error term ($f$). We applied this modeling approach to the $\gamma$-ray light curves of five blazars: PKS 1424-41, S2 0109+22, PKS 0244-470, PKS 0405-385, and PKS 0208-512. The MCMC simulation was run for 50,000 iterations, with the first 5,000 steps discarded as burn-in. The posterior distributions of the model parameters were then derived from the remaining samples. The fitted light curves and corresponding posterior distributions of the parameters are presented in Figures~\ref{Fig-mcmc_lc}, ~\ref{Fig-MCMC_LC_fitting}, and listed in Table~\ref{tab:SMBBH_posterior}. The derived QPO timescale is significantly shorter than the actual physical period due to the light-travel time effect \citep{rieger2004geometrical}. The QPO timescale is related as $P_d = \Gamma^2 P_{int}$, where $\Gamma$ is the bulk Lorentz factor. Based on the corrected QPO timescale, the mass of the primary black hole is given as

\begin{equation}
    M \simeq P_{d,\ yr}^{8/5} \ R^{3/5} \ 10^6 \ M_{\odot}
    \label{eq:primary_bh_mass}
\end{equation}

where $P_{d, \ yr}$ is the corrected physical timescale in units of years, R denotes the mass ratio of the primary black hole to the secondary black hole, $R = \frac{M}{m}$. In this study, we adopted a mass ratio of R$\sim 1$ to estimate the masses of the primary black holes in our sample. Using this assumption, the inferred black hole masses are approximately $1.3\times10^9 \ M_{\odot}$ for PKS 1424-41, $1.8\times10^{10} \ M_{\odot}$ for S2 0109+22, $9.5\times10^8 \ M_{\odot}$ for PKS 0244-470, $3.5\times10^9 \ M_{\odot}$ for PKS 0405-385, and $4.4\times10^{10} \ M_{\odot}$ for PKS 0208-512. These mass estimates were obtained using Equation~\ref{eq:primary_bh_mass}, incorporating the Lorentz factor ($\Gamma$) derived individually for each source from the modeling of the $\gamma$-ray light curves under the SMBBH framework, rather than assuming a fixed $\Gamma$. For S2 0109+22 and PKS 0208-512, the estimated black hole masses are somewhat higher than those reported in previous studies but remain consistent within the uncertainties associated with the $\Gamma$ parameter. Given the long timescales of the detected QPOs, the presence of a close supermassive binary black hole system, where one black hole launches a relativistic jet-provides a compelling scenario to explain both the observed QPO behavior and its physical origin.

\begin{table*}
\setlength{\extrarowheight}{10pt}
\setlength{\tabcolsep}{12pt}
\centering
\begin{threeparttable}
\caption{Summary of the posterior probability distributions of blazars $\gamma$-ray light curves modeling under the SMBBH scenario.}
\label{tab:SMBBH_posterior}
\begin{tabular}{c c c c c c}
\hline
\hline
Source  & \multicolumn{5}{c}{Parameter} \\
\cline{2-6}
 & $\Gamma$ & $\psi \ (deg)$ & $P \ (days)$ & $F_0 \ (\mathrm{ph \ cm^{-2} \ s^{-1}})$ & logf\\
(1) & (2) & (3) & (4) & (5) & (6)\\
[+2pt]
\hline
PKS 1424-41 & $15.35_{-5.57}^{+8.87}$ & $1.34_{-0.32}^{+0.39}$ & $358.01_{-2.21}^{+2.27}$ & $6.5_{-3.58}^{+7.25} \ \times \ 10^{-9}$ & $-0.47_{-0.05}^{+0.05}$ \\
S2 0109+22 & $19.96_{-12.41}^{+6.49}$ & $1.80_{-0.42}^{+0.36}$ & $649.52_{-0.09}^{+0.09}$ & $1.3_{-0.87}^{+3.65} \ \times \ 10^{-9}$ & $-0.86_{-0.06}^{+0.06}$\\
PKS 0244-470 & $16.76_{-10.71}^{+8.09}$ & $1.20_{-0.32}^{+0.50}$ & $245.91_{-1.39}^{+1.41}$ & $3.64_{-3.52}^{+4.24} \ \times \ 10^{-8}$ & $-0.60_{-0.08}^{+0.08}$\\
PKS 0405-385 & $12.02_{-5.37}^{+3.69}$ & $1.10_{-0.12}^{+0.28}$ & $1006.31_{-0.37}^{+0.36}$ & $4.29_{-1.98}^{+4.85} \ \times \ 10^{-10}$ & $-0.84_{-0.06}^{+0.06}$\\
PKS 0208-512 & $26.22_{-8.95}^{+8.53}$ & $0.80_{-0.13}^{+0.14}$ & $871.76_{-0.11}^{+0.11}$ & $1.01_{-0.46}^{+1.34} \ \times \ 10^{-9}$ & $-0.75_{-0.04}^{+0.05}$\\
[+5pt]
\hline
\end{tabular}
\begin{tablenotes} 
\small
\item Note: Details of posterior probability distributions of $\gamma$-ray light curves modeling within SMBBH scenario. Column (1): Lorentz factor; Column (2): angle between the line of sight and spin axis; Column (3): QPO period; Column (4): baseline flux; Column (5): fractional error term.\par

$\zeta^*: $ adopted as $5^{\circ}$ in the light curve modeling. 

\end{tablenotes}
\end{threeparttable}
\end{table*}

\section{Discussion and conclusion}\label{sec:disscusion}
In this study, we investigated quasi-periodic oscillations (QPOs) in the $\gamma$-ray emission from a sample of blazars over a 15-year period (MJD 54686–60769). By applying multiple time-series analysis techniques—including the Lomb-Scargle Periodogram (LSP) and REDFIT—we identified both long-term and transient QPO features with timescales ranging from several months to a few years. To evaluate the local significance of these signals, we modeled the light curves using a damped random walk (DRW) process. The LSP of PKS 0736+01 reveals a dominant peak with a significance exceeding $4\sigma$, while significant peaks above $3\sigma$ are detected for PKS 1424-41, S2 0109+22, and PKS 0244-470. The remaining sources also exhibit QPO signatures with significance close to $3\sigma$. These results are further supported by REDFIT analysis, which confirms the presence of QPOs at confidence levels exceeding 99$\%$ across all sources. Notably, PKS 0736+01 displays approximately four cycles of QPOs in its $\gamma$-ray light curve, with a periodicity of around 4 years-placing it among the few blazars showing year-scale QPOs, which can be an ideal candidate for future studies. Uncertainties in the observed QPO periods were quantified using the half-width at half-maximum (HWHM) from Gaussian fits to the dominant peaks. \par

To interpret the observed quasi-periodic oscillations (QPOs) in the $\gamma$-ray light curves of blazars in our sample, several physical scenarios have been considered. These include the presence of a supermassive binary black hole (SMBBH) system, jet precession or helical motion, and accretion disk instabilities. For instance, blazar PKS 0035-252 exhibits a QPO with a timescale of approximately 112 days—a month-scale periodicity that can plausibly be attributed to a magnetized plasma blob spiraling downward along a helical jet trajectory. A similar model was tested for PKS 0244-470; however, previous studies \citep{rieger2004geometrical, mohan2015kinematics} argue that such jet-induced helical motion cannot account for QPOs with timescales exceeding a few months. To explain QPOs with longer timescales, we employed an SMBBH framework in which one of the black holes launches a relativistic jet. Within this model, we utilized a Markov Chain Monte Carlo (MCMC) approach to estimate key parameters such as the QPO period, Lorentz factor ($\Gamma$), and viewing angle ($\psi$). Our analysis supports the hypothesis that a single relativistic jet, modulated by the orbital dynamics of a close SMBBH system, can produce the longer-period oscillations observed in blazars such as PKS 0736+01, PKS 1424-41, S2 0109+22, PKS 0244-470, PKS 0405-385, and PKS 0208-512. This modeling approach not only provides a physically motivated explanation for long-timescale QPOs but also offers an effective alternative framework for identifying periodic signals in astrophysical time series data.


\newcommand{\aap}{Astronomy \& Astrophysics}
\newcommand{\apj}{The Astrophysical Journal}
\newcommand{\ssr}{Space Science Reviews}
\newcommand{\mnras}{Monthly Notices of the Royal Astronomical Society}
\newcommand{\apjl}{The Astrophysical Journal Letters}
\newcommand{\pasp}{Publications of the Astronomical Society of the Pacific}




\section{Acknowledgements}
A. Sharma is grateful to Prof. Sakuntala Chatterjee at S.N. Bose National Centre for Basic Sciences for providing the necessary support to conduct this research.



\paragraph{\textbf{Data Availability Statement}}
This research utilizes publicly available data of blazars in our sample obtained from the Fermi-LAT data server provided by NASA Goddard Space Flight Center (GSFC): \url{https://fermi.gsfc.nasa.gov/ssc/data/access/}.

\bibliographystyle{elsarticle-harv}

\newpage
\bibliography{example}




\appendix
\section{Gamma-ray flux distribution}\label{sec:flux_distribution}

We analyzed the weekly binned $\gamma$-ray flux distributions for all blazars in our sample, selecting only those data points with a test statistic $TS \ge 9$ and $F_i > \sigma_i$, thereby excluding upper limits. The logarithm of the flux values was modeled using a Gaussian distribution. In all cases, the resulting $p$-values were less than 0.01, indicating significant deviation from a normal distribution when considering the flux in linear space. However, when analyzed in logarithmic space, the flux distributions are well described by a Gaussian, suggesting that the intrinsic $\gamma$-ray flux variability of all sources in our sample follows a log-normal distribution, Figure~\ref{Fig-LC_FLUX_distribution}.

\section{DRW modeling}\label{sec:appendex_DRW}
The observed DRW model parameters are listed in Table~\ref{tab:DRW}.
\begin{figure*}
    \centering
    \includegraphics[width=0.3\textwidth]{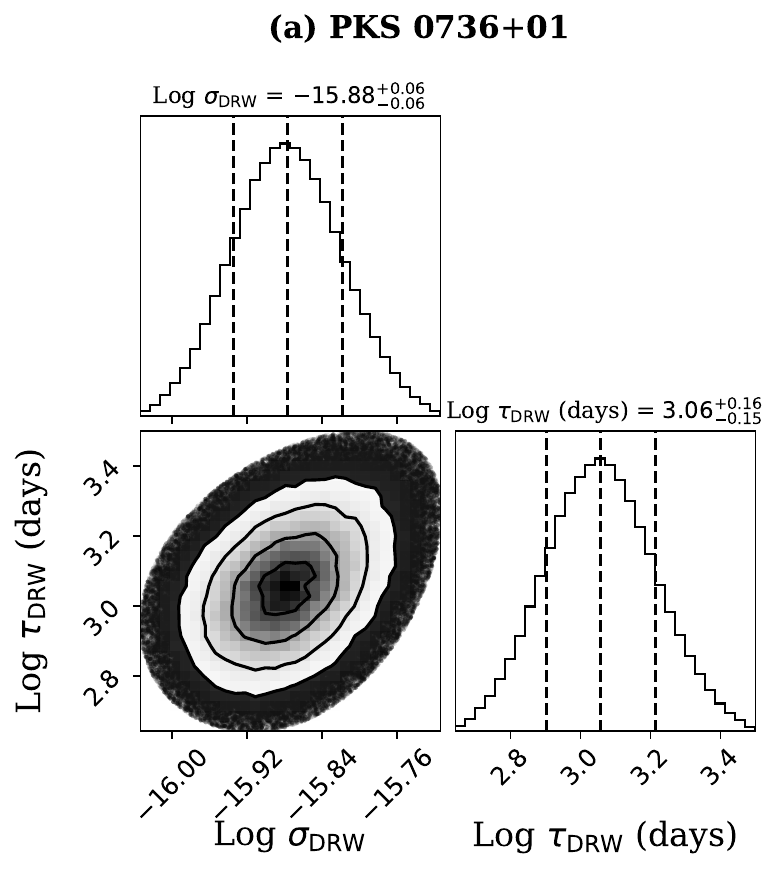} \hspace{1pt}
    \includegraphics[width=0.3\textwidth]{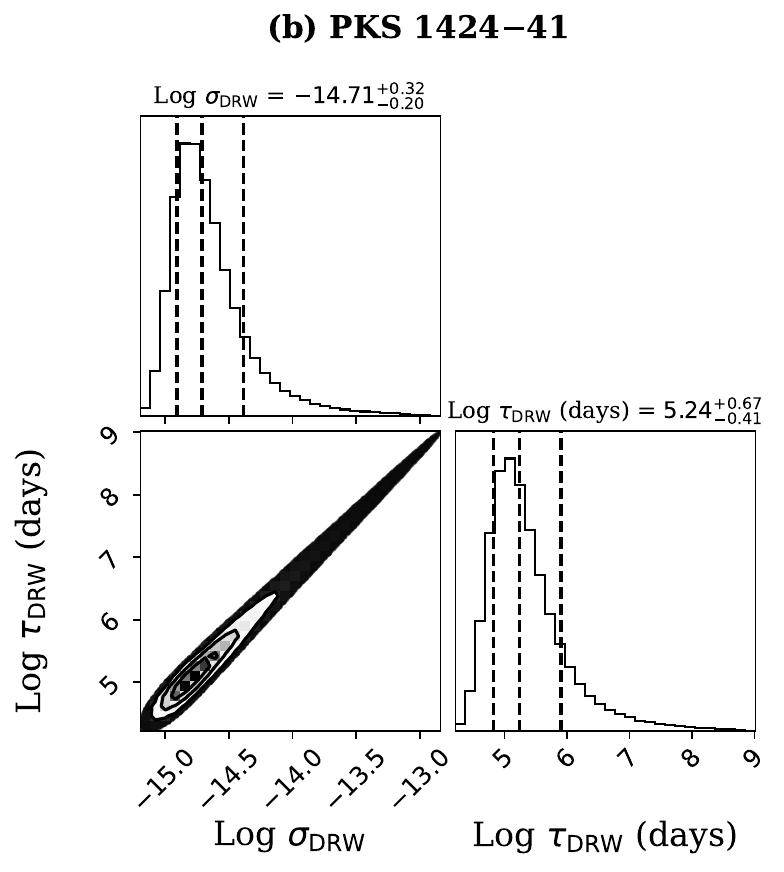} \hspace{1pt}
    \includegraphics[width=0.3\textwidth]{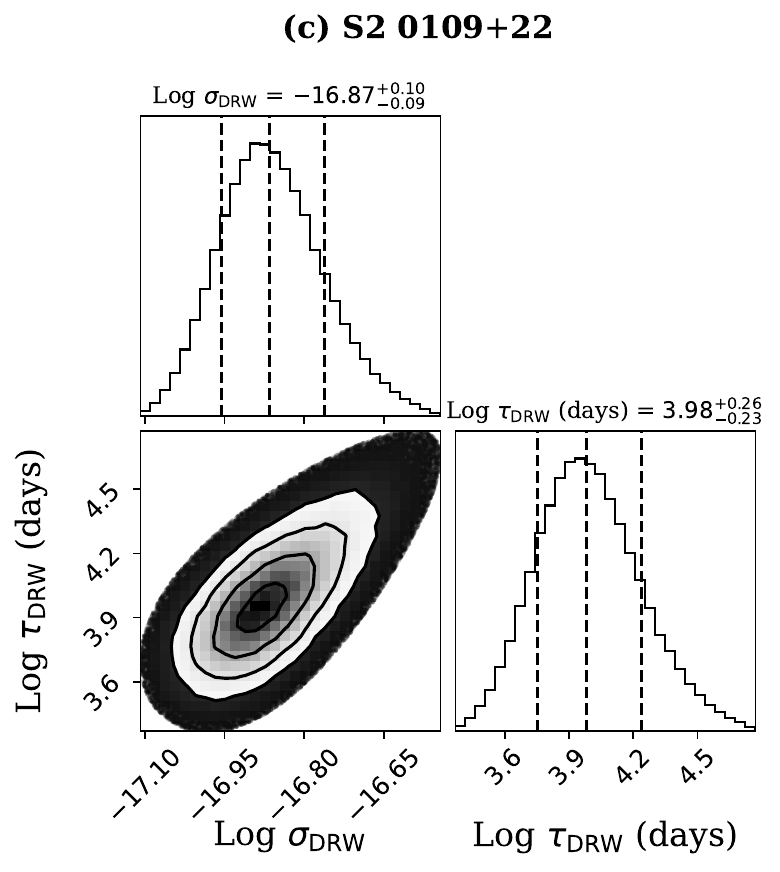} \vspace{1pt}
    \includegraphics[width=0.3\textwidth]{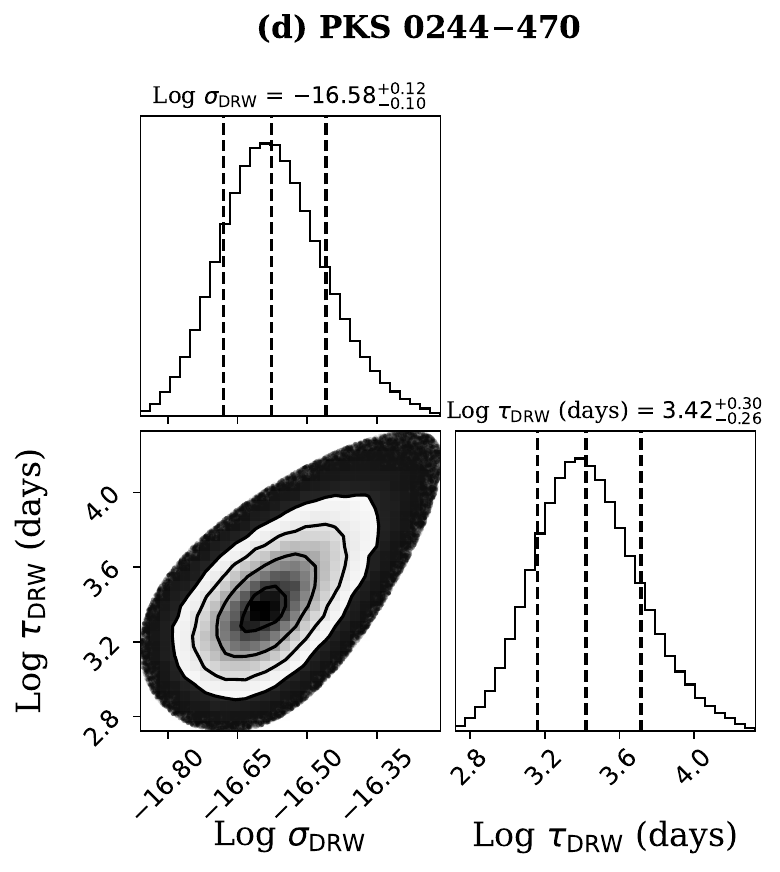} \hspace{1pt}
    \includegraphics[width=0.3\textwidth]{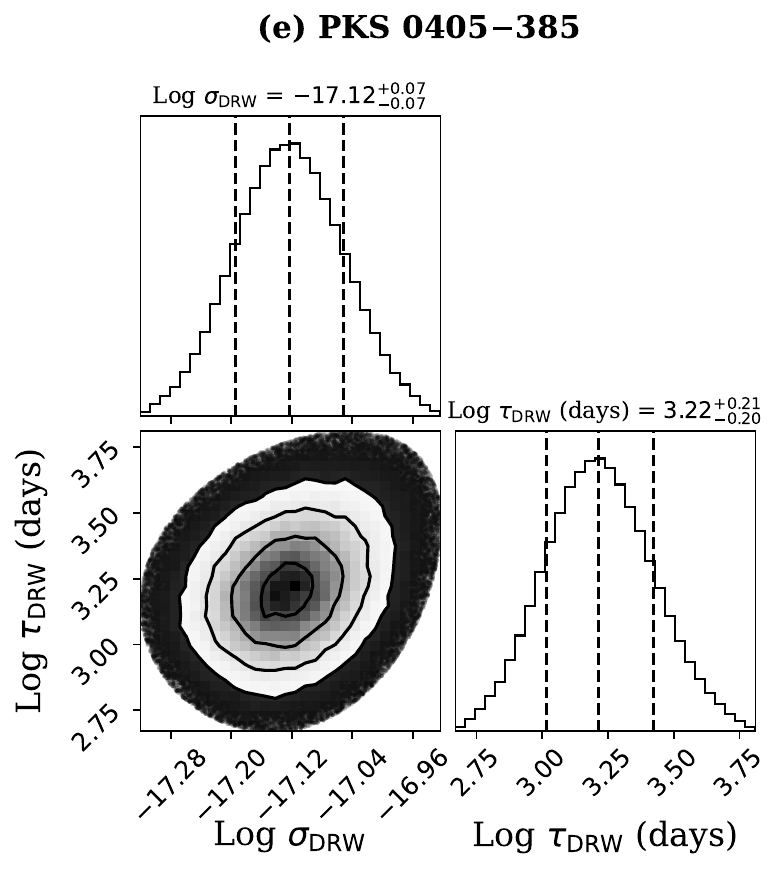} \hspace{1pt} 
    \includegraphics[width=0.3\textwidth]{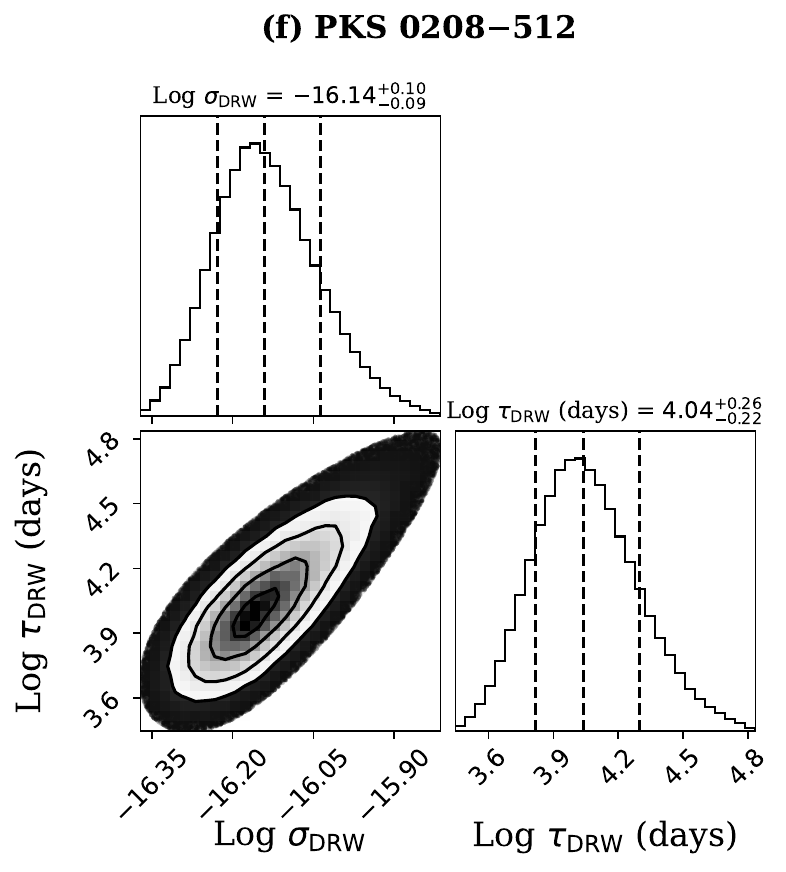} \vspace{1pt}
    \includegraphics[width=0.3\textwidth]{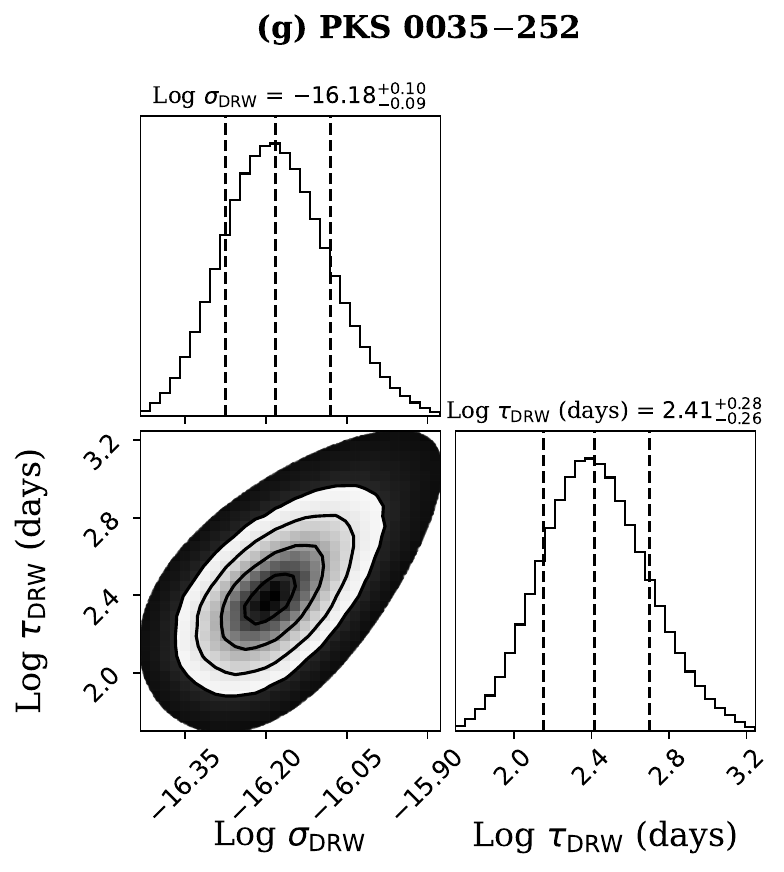}
    \caption{The figure displays the posterior distributions of the DRW model parameters, obtained from the modeling of the $\gamma$-ray light curve of blazars in our sample. }
    \label{Fig-DRWCorner}    
\end{figure*}

\begin{table}
\setlength{\extrarowheight}{7pt}
\setlength{\tabcolsep}{8pt}
\centering
\caption{DRW model parameters from the modeling of $\gamma$-ray light curves of blazars listed in Table~\ref{tab:source_sample}.}

\begin{tabular}{c c c}
\hline
\hline
Source  & $\rm{Log} \ \sigma_{DRW}$ & $\rm{Log} \ \tau_{DRW} \ (\rm{days})$ \\
(1) & (2) & (3) \\
[+2pt]
\hline
PKS 0736+01 &  -15.88$_{-0.06}^{+0.06}$ & 3.06$_{-0.15}^{+0.16}$ \\ 
PKS 1424-41 &  -14.71$_{-0.20}^{+0.32}$ & 5.24$_{-0.41}^{+0.67}$ \\
PKS 1424-41 &  -16.87$_{-0.09}^{+010}$ & 3.98$_{-0.23}^{+0.26}$ \\
S2 0109+22 &  -14.71$_{-0.20}^{+0.32}$ & 5.24$_{-0.41}^{+0.67}$ \\
PKS 0244-470 &  -16.58$_{-0.10}^{+0.12}$ & 3.42$_{-0.26}^{+0.30}$ \\
PKS 0405-385 &  -17.12$_{-0.07}^{+0.07}$ & 3.22$_{-0.20}^{+0.21}$ \\
PKS 0208-512 &  -16.14$_{-0.09}^{+0.10}$ & 4.04$_{-0.22}^{+0.26}$ \\
PKS 0035-252 &  -16.18$_{-0.09}^{+0.10}$ & 2.41$_{-0.26}^{+0.28}$ \\
[+5pt]
\hline
\end{tabular}

\label{tab:DRW}
\end{table}


\end{document}